\documentclass[twocolumn,pre,floatfix]{revtex4}
\usepackage{psfrag,epsfig,amsfonts,amssymb,amsmath,wasysym,bm}
\usepackage{dcolumn}
\usepackage{bbm,psfrag}

\usepackage[normalem]{ulem}
\usepackage{color}

\newcommand{\nn}{{\cal N}}
\newcommand{\pu}{\tr\{\rho^2\}}
\newcommand{\deff}{d_{\mathrm{eff}}}
\newcommand{\oo}{'}
\newcommand{\ak}{\alpha}

\newcommand{\id}{\ensuremath{\mathbbm{1}}}
\newcommand{\rhomic}{\rho_{\mathrm{mc}}}

\newcommand{\lsim}{\mathrel{\hbox{\rlap{\lower.55ex \hbox{$\sim$}} \kern-.3em \raise.4ex \hbox{$<$}}}}
\newcommand{\gsim}{\mathrel{\hbox{\rlap{\lower.55ex \hbox{$\sim$}} \kern-.3em \raise.4ex \hbox{$>$}}}}

\newcommand{\hr}{{\cal H}}
\newcommand{\ord}{{\cal O}}

\newcommand{\tr}{\mbox{Tr}}

\newcommand{\db}{\Delta_{\! B}}

\providecommand{\norm}[1]{\|#1\|}

\newcommand{\RR}{{\mathbb R}}
\newcommand{\CC}{{\mathbb C}}

\begin{document}

\title{Why are macroscopic experiments reproducible\,?
\\
Imitating the behavior of an ensemble by single pure states
}

\author{Peter Reimann}
 \email{reimann@physik.uni-bielefeld.de}
 \affiliation{Fakult\"at f\"ur Physik, 
Universit\"at Bielefeld, 
33615 Bielefeld, Germany}

\author{Jochen Gemmer}
 \email{jgemmer@uos.de}
 \affiliation{Department of Physics, University of Osnabr\"uck, D-49069 Osnabr\"uck, Germany}

\begin{abstract}
Evidently, physical experiments are practically reproducible even 
though the fully identical preparation of 
initial state wave functions is often far beyond experimental possibilities. 
It is thus natural to explore if and in which sense specific,
uncontrollable features of initial wave functions are irrelevant for 
the observable course of an experiment. 
To this end we define ensembles of pure states
which are then shown to generate extremely similar non-equilibrium 
dynamics of the expectation values of practically all standard observables. 
The  ensembles are constructed to comply with some reduced, 
coarse {\it a priori} information on the state of the system, like, 
e.g. a few specific expectation values, etc. 
However, different types of ensembles with different additional properties 
are possible. We discuss some of them.
\end{abstract}

\maketitle

\section{Introduction}
\label{s0}
Appreciating simplicity and surprises
on a fundamental level
were among the main things we could learn from 
our unforgettable friend and teacher Chris
Van den Broeck.
One of us (P.R.) had the great privilege to spend
more than two years with him as a postdoc
during the heyday of exploring 
neural networks in terms of 
statistical mechanics.
In this context, he 
happily infected
all of us with his excitement
about the 
intriguing
geometrical 
properties of high-dimensional 
Euclidean spaces \cite{eng01}, 
such as the 
well-known
peculiarities of a simple sphere:
Almost its entire volume 
is concentrated in an exceedingly thin surface layer,
which,  in turn, exhibits an extreme 
concentration around a very narrow 
``equatorial belt'' or any other ``great circle'', 
and hence also around the intersection 
of two such circles,
and so on.

Some ten years later, a related 
surprise in high-dimensional Hilbert 
spaces attracted a lot of interest under 
the labels of ``canonical typicality'' 
\cite{gol06}
and ``concentration of measure'' 
\cite{pop06}.
Namely, the rediscovery \cite{llo88}
that the overwhelming majority 
of randomly sampled 
wave functions
(vectors on the unit sphere) 
exhibit almost identical expectation values.
Obviously, this was exactly the type 
of thing which 
fascinated Chris,
in particular since he immediately 
recognized its far reaching physical
implications: All thermal fluctuation
effects could, in principle, be ascribed
to the purely quantum mechanical 
fluctuations of a pure state \cite{llo88}.
Yet he remained 
unhappy with the rather unintuitive 
``Levy lemma'', exploited in \cite{pop06} 
to derive those results, but only 
for a few days;
the following email from March 2007,
entitled ``Levy lemma obvious'', reminds
us once more of his inimitable style:
``Yes it is obvious, I have understood it while 
sitting in the bath, like the great Archimedes ...
Consider the (closed) line of points $x$ on the 
sphere for which $f(x)$ takes on the
median value. This line divides by definition 
the sphere in two equal parts.
This line is at least in length like an equator. 
Any small neighborhood of the
equator basically covers the whole sphere 
in high dimensions. This is then 
{\em a fortiori} so for the neighborhood of 
the median line, which is due to the Lipschitz 
condition also the values of $f$ close to 
the median. That is it.''

At the focus of our present work is the
question of how far one can go in extending
those originally {\em non-dynamical} typicality
concepts \cite{gol06,pop06,llo88}
into the realm of 
dynamical typicality \cite{bar09},
and how they can be efficiently
utilized for practical 
(analytical and numerical) purposes.

\section{Main issues and results}
\label{s1}
\subsection{Core message}
\label{s11}
It is a basic scientific principle and empirical 
fact that repeating a macroscopic experiment 
at different times or in different labs 
results in reproducible observations.
On the other hand, some small 
differences in the microscopic initial 
conditions are practically unavoidable 
and will subsequently grow 
very fast due to the 
generically chaotic dynamics, so that 
the similarity of the final results 
becomes difficult to explain on a 
fundamental theoretical level.

Leaving aside the technical details, 
the main result of our present 
explorations provides the following 
general picture of how the above 
antinomy can be resolved within
the framework of elementary quantum 
mechanics:
Taking for granted that the initial 
system energy is known (is reproducible) 
up to an uncertainty which is small 
on macroscopic but large on 
microscopic scales, the overwhelming
majority of initial (pure) states
$|\psi\rangle$ 
which are compatible with this 
information 
(energy)
yields very similar expectation 
values $\langle\psi|O|\psi\rangle$
for any given observable $O$,
which in turn is very close to the
expectation value $\tr\{\rhomic O\}$,
where $\rhomic$ is the textbook 
microcanonical ensemble for the 
given system energy. 
In fact, this is the essence of 
the well-known (canonical) typicality 
and concentration of measure phenomena 
\cite{llo88,gol06,pop06,sug07,rei07,gem09,sug12,tas16},
which were mentioned already in Sec. \ref{s0}.

To characterize a {\em non-equilibrium}
initial condition, additional 
information (besides the system 
energy) is thus indispensable.
In the simplest case, it is known
that the expectation value 
$\langle\psi|A_1|\psi\rangle$ 
of some observable $A_1$
(e.g. some macroscopic system
property) notably deviates from
the corresponding equilibrium value 
$\tr\{\rhomic A_1\}$, and 
instead is (approximately) 
equal to some other value $\ak_1$.
Similarly as before, 
we will show
that the vast majority of initial states
$|\psi\rangle$ which (approximately)
reproduce the given system energy
{\em and} the non-equilibrium 
expectation value of $A_1$,
yield very similar expectation 
values $\langle\psi|O|\psi\rangle$
also for any other observable $O$.
If there is yet another observable $A_2$,
whose expectation value $\ak_2$ is 
known to notably deviate from the 
latter typical value, then there 
will be yet another ensemble
of $|\psi\rangle$'s, which 
is compatible with the energy {\em and}
the given $\ak_1$ and $\ak_2$ values,
and which again yields predominantly very 
similar $\langle\psi|O|\psi\rangle$ 
values, and so on.

In many experimental situations it 
seems reasonable to expect that
there exists a finite, relatively 
small number $K$ of such steps,
after which {\em all} relevant
non-equilibrium information will 
have been satisfactorily taken into account.
(Whether or not one {\em explicitly} 
knows all those $A_k$'s and $\ak_k$'s
does not matter.)
Furthermore, one expects that
the ``true'' initial state 
$|\psi\rangle$ now belongs to the concomitant
vast majority of states, which all exhibit very similar 
$\langle\psi|O|\psi\rangle$ 
values for $O=A_k$, $k=1,...,K$,
but also for practically any other physically relevant 
observable $O$.

By employing the Heisenberg picture of
quantum mechanics, one can conclude
\cite{rei07} that also the expectation 
value of $O$ at any later instant of 
time will be very similar for most of those initial 
states $|\psi\rangle$, a property for which 
the term {\em dynamical typicality} has been 
coined in Ref. \cite{bar09}.

In other words, we propose a new
way of classifying non-equilibrium 
initial states of many-body 
systems (at a given energy), 
namely according to the 
pertinent set of observables $A_k$ 
and their expectation values $\ak_k$.
Exactly as in an experiment, for
the vast majority of initial states
which are compatible with this 
information, all measurable properties 
of the system both at the initial and 
also at any later time point 
turn out to be very similar 
(the differences are either below any 
reasonable resolution limit or their 
probability is negligibly small).
All the further details, which 
inevitably must remain unknown
and irreproducible in an experiment, 
are thus indeed ``irrelevant'' 
in our theory as well.

\subsection{Ramifications}
\label{s12}
If the considered system is isolated
from the rest of the world, then
it will usually exhibit a temporal
relaxation towards thermal equilibrium.
But it should be emphasized that also
systems which do not thermalize and 
even non-autonomous systems with
explicitly time-dependent Hamiltonians
are still admitted by the general 
framework of our present work.

On the other hand, standard quantum 
mechanics cannot directly deal with 
open systems, 
interacting and entangled with some 
environment.
In such a case, we therefore tacitly 
consider, as usual, the entire 
system-plus-environment compound 
as the actual (super-)system of interest,
which then can again be treated as
isolated from the rest of the world.

So far, the entire discussion was 
conducted in terms of expectation values,
i.e., ensemble averaged measurement outcomes
when repeating the same experiment many 
times.
On the other hand, due to the fundamentally
probabilistic nature of quantum mechanics,
even when the considered setup 
(system state and observable) is
always exactly identical (which
is not feasible in a real experiment 
but is an admissible Gedankenexperiment),
the actual outcomes of the single 
measurements in general give rise
to a non-trivial statistical
distribution.
This issue is of particular relevance for 
microscopic observables (e.g., the Brownian 
motion of a small particle, or the velocity 
distribution of one molecule), where
often not only the average but rather
the entire measurement outcome
distribution may be of interest.
The corresponding generalization of our
approach is straightforward, 
predicting that even the entire statistics 
of the different possible measurement outcomes 
will be almost the same for practically 
all initial states with common 
$\langle\psi|A_k|\psi\rangle$ 
values.
The reason is that with any given observable
$O$, also the higher moments $O^2$, $O^3$, ...,
or equivalently, the projectors onto
any eigenspace of $O$, are 
admissible observables with very 
similar expectation values for 
most $|\psi\rangle$'s.

Note that in reality, the actually
measurable fluctuations of a microscopic 
observable (e.g. the position of a
fluorescent 
molecule
in a fluid)
are unavoidably ``contaminated'' 
with differences in the initial states
when repeating the ``same'' experiment.
Similarly as in the original (canonical)
typicality investigations 
\cite{llo88,gol06,pop06,sug07,rei07,gem09,sug12,tas16}, 
a conceptual key
aspect of our approach is that those
microscopic fluctuations would actually
remain largely the same even if the initial
states were strictly identical in every
repetition of the experiment.

In the same vein, our approach
allows one to (approximately)
preset not only the initial expectation 
values of certain observables $A_k$ but 
even the entire statistics of their
different possible measurement outcomes.
For instance, one may 
impose some expectation value 
$\ak_1$ for an observable $A_1$
and then choose as a second 
observable $A_2:=(A_1-\ak_1)^2$
together with a possibly quite
small (positive) value of $\ak_2$.
In this way, not only the mean
value of the (random) measurement 
outcomes of $A_1$, but also the 
variance can be prescribed, 
and hence will be closely reproduced 
with high probability by an arbitrary 
but fixed sample of the 
corresponding ensemble of pure 
states $|\psi\rangle$.
Analogously, further details of the 
measurement outcome statistics can be 
imposed via the third and higher 
moments of $A_1$, or equivalently,
via the projectors onto the 
eigenspaces of $A_1$ and the
corresponding projection 
probabilities \cite{bal18}.

Besides the so far discussed lack 
of knowledge about the exact system
state $|\psi\rangle$, also the Hamiltonian, 
which governs the time-evolution and even
the pertinent Hilbert space may not be
exactly known and/or may vary upon 
repeating the ``same'' experiment.
However, this issue goes beyond the 
scope of our present work.

\section{Dynamical Typicality}
\label{s2}
We consider a Hilbert space $\hr$,
spanned by some orthonormal basis
$\{|\chi_n\rangle\}_{n=1}^N$, whose
dimension $N$ may be either large 
but finite or infinite.
For instance, every $|\chi_n\rangle$ may 
be an eigenvector of some Hamiltonian 
$H$, and $\hr$ may be either
an energy shell (spanned by a 
finite subset of all eigenvectors 
of $H$) or the full Hilbert space 
(spanned by all eigenvectors of 
$H$), see also Sec. \ref{s4} below.

As in Sec. \ref{s1}, we mainly have 
in mind macroscopic (or ``many-body'') 
systems, for which the dimension of
a typical microcanonical energy shell 
is well-know to be exponentially 
large in the number of microscopic
degrees of freedom, while the full 
Hilbert space will be of even 
larger and possibly infinite dimensionality.

In terms of the above introduced basis 
$\{|\chi_n\rangle\}_{n=1}^N$ of $\hr$,
we define an ensemble of random 
vectors $|\phi\rangle\in\hr$ via
\begin{eqnarray}
|\phi\rangle=
\sum_{n=1}^N R\, c_n \, |\chi_n\rangle
\ ,
\label{1}
\end{eqnarray}
where the $c_n$ are complex numbers,
whose real and imaginary parts are 
given by independent, Gaussian distributed 
random variables of mean zero and 
variance $1/2$.
Moreover, $R$ is a linear operator on $\hr$,
which for the moment may still be 
arbitrary apart  from the requirement 
that the positive semidefinite 
Hermitian operator
\begin{eqnarray}
\rho:=R\,R^\dagger 
\label{2}
\end{eqnarray}
is of finite trace.
Obviously, the latter requirement is
non-trivial only if $N$ is infinite and
guarantees, as we will see in a moment, 
that  the random vectors in (\ref{1}) are 
of finite norm (with probability one).
For the rest, the random vectors
(\ref{1}) are not yet normalized,
rather their norm is itself a random 
variable.
Moreover, it will turn out that the 
finite trace of $\rho$ can be set to unity 
without any loss of generality later on, 
i.e., we require that
\begin{eqnarray}
\tr\{\rho\}=1 \ .
\label{3}
\end{eqnarray}
As anticipated by the notation,
$\rho$ is thus a well defined
density operator (Hermitian, positive 
semidefinite, and of unit 
trace).

Note that $R$ on the right hand side
of (\ref{1}) could be readily taken out 
of the sum if $N$ is finite, but for 
infinite $N$ the remaining sum (without $R$)
would not be well defined.

A key property of the random vector ensemble 
in (\ref{1}) is its invariance under arbitrary 
unitary transformations of the basis 
$|\chi_n\rangle$ of $\hr$
(all statistical properties 
remain unchanged).
In other words, the basis can be 
chosen arbitrarily.
This is of particular interest when numerically
sampling random vectors according to (\ref{1})
on the computer: 
For instance, any single-particle product basis 
will do in the case of a many-body system.
Furthermore, focusing on the special choice 
$R=\id/\sqrt{N}$
(which  is only possible for $N<\infty$,
and where $\id$ is the identity on $\hr$)
we can conclude that after normalization 
of each random vector in (\ref{1}), all those 
normalized vectors will be equally 
likely (uniformly distributed) 
in a very natural sense.

Next we consider an arbitrary Hermitian
operator $B:\hr\to\hr$. 
If the dimension $N$ of $\hr$ is infinite, 
the spectrum of $B$ is furthermore
required to be bounded.
In other words, the operator 
norm $\norm{B}$ (largest eigenvalue in 
modulus) is required to be finite.
Since only $B$'s which model some
observable will actually be of interest 
later on, and since the measurement
range of any real observable is finite, 
the latter requirement does not amount
to any significant loss of generality.

Denoting the average over the random 
vector ensemble from (\ref{1}) by an
overbar, one can readily show that
\begin{eqnarray}
\bar B & := & \overline{\langle\phi|B|\phi\rangle}
= \tr\{\rho  B\}
\ ,
\label{4}
\\
\sigma_B^2 & := &
\overline{
\left(\langle\phi|B|\phi\rangle
-
\bar B
\right)^2
}
=
 \tr\{(\rho  B)^2\} 
\label{5}
\end{eqnarray}
by means of the following line of reasoning:
In order to verify (\ref{4}), we utilize
(\ref{1}) and the abbreviation
$B':=R^\dagger B R$ to obtain
\begin{eqnarray}
\bar B= \sum_{m,n=1}^N \overline{c_m^\ast c_n}\,
\langle\chi_m|B'|\chi_n\rangle 
\ .
\label{6}
\end{eqnarray}
Since the real and imaginary parts of the $c_n$'s 
are independent, Gaussian distributed 
random variables of mean zero and 
variance $1/2$ (see below (\ref{1})), 
it follows that
$\overline{c_m^\ast c_n}=\delta_{mn}$.
Eq. (\ref{6}) thus takes the form
\begin{eqnarray}
\bar B= \sum_{n=1}^N \, \langle\chi_n |B'|\chi_n\rangle
=\tr\{B'\} \ .
\label{7}
\end{eqnarray}
Exploiting the definition of $B'$, the cyclic 
invariance of the trace,  and the definition 
(\ref{2}), one recovers (\ref{4}).
Upon verifying and utilizing that
$\overline{c^\ast_jc_kc^\ast_mc_n}=
(\delta_{jk}\delta_{mn}+\delta_{jn}\delta_{km})$,
a similar calculation yields (\ref{5})
(analogous calculations can 
also be found, e.g., in Refs. 
\cite{rei07,gem09}).

Considering $\tr\{C_1^\dagger C_2\}$
as a scalar product between two arbitrary
linear (but not necessarily Hermitian) 
operators $C_{1,2}:\hr\to\hr$, 
the Cauchy-Schwarz
inequality takes the form
$|\tr\{C_1^\dagger C_2\}|^2\leq
\tr\{C_1^\dagger C_1\}\tr\{C_2^\dagger C_2\}$.
Choosing $C_1=B\rho $ and $C_2=\rho B$,
it follows that $\tr\{(\rho B)^2\}=\tr\{C_1^\dagger C_2\}$
and that both $\tr\{C_1^\dagger C_1\}$
and $\tr\{C_2^\dagger C_2\}$
can be rewritten as $\tr\{\rho ^2B^2\}$.
With (\ref{5}) we thus obtain
\begin{eqnarray}
\sigma_B^2 \leq \tr\{\rho ^2 B^2\}
\ .
\label{8}
\end{eqnarray}
Evaluating the trace by means of the 
eigenbasis of $B$, one finds
that 
\begin{eqnarray}
\sigma_B^2 \leq \norm{B}^2\,  \pu 
\ ,
\label{9}
\end{eqnarray}
where $\norm{B}$ is the operator norm 
introduced above (\ref{4}), and where
$\pu$
is the purity of the density 
operator $\rho$ from (\ref{2}).

In the following, we restrict ourselves to
operators $R$ in (\ref{1}), (\ref{2}) 
with the property that 
\begin{eqnarray}
\pu  \ll 1
\label{10}
\end{eqnarray}
for reasons which will become obvious 
immediately.

Choosing for $B$ the identity operator, 
it follows from (\ref{3})-(\ref{5}) that
\begin{eqnarray}
& & \overline{\langle\phi|\phi\rangle} = 1
\ ,
\label{11}
\\
& & 
\overline{
\left(\langle\phi|\phi\rangle
- 1 \right)^2
}
=\pu 
\ .
\label{12}
\end{eqnarray}
With (\ref{10}) we can conclude that the 
overwhelming majority of all vectors 
$|\phi\rangle$ in (\ref{1})
have norms very close to unity.

Note that the variance $\sigma_B^2$ in (\ref{5})
vanishes for $B=0$, and is generically 
positive for $B=\id$
according to (\ref{12}).
More generally, $\sigma_B^2$ is {\em not}
invariant when replacing $B$ in (\ref{5})
by $B+b\,\id$ with $b\in\RR$.
The reason for this quite unusual behavior
is that the norm of the vectors in (\ref{1}) 
is not fixed but rather exhibits random 
fluctuations.

Next, we consider any given $|\phi\rangle$ in
(\ref{1}) as an initial state at time $t=0$, 
which then evolves according to the 
time-dependent Schr\"odinger equation 
into the state $|\phi(t)\rangle=U_t|\phi\rangle$
after a time $t$ has elapsed,
where $U_{t}$ is the 
unitary quantum-mechanical time-evolution operator.
For a time independent Hamiltonian $H$, 
the propagator $U_{t}$ takes 
the simple form $\exp\{-iHt/\hbar\}$, 
but in full generality, also any 
explicitly time-dependent 
Hamiltonian $H(t)$ is admitted
(see also Appendix \ref{app1}).
In particular, the system is not required to 
exhibit equilibration or thermalization
in the long-time limit.

Going over from the Schr\"odinger to the 
Heisenberg picture of quantum mechanics,
the corresponding expectation value for an arbitrary
observable $O$ (of finite operator norm, see above)
at an arbitrary but fixed time $t$ 
thus follows as
\begin{eqnarray}
& & 
\langle \phi(t)|O|\phi(t)\rangle
=\langle\phi|O_t|\phi\rangle
\ ,
\label{13}
\\
& & O_t := U_t^\dagger O\, U_t \ ,
\label{14}
\end{eqnarray}
where we ignore for the time being that
the vector $|\phi(t)\rangle$ may not be 
normalized.
Choosing for $B$ the operator $O_t$ 
we can conclude from (\ref{4}), (\ref{5}), (\ref{9}),
and (\ref{10}) that the vast majority 
of all vectors 
$|\phi\rangle$ in (\ref{1}) exhibit 
expectation values very close to
the mean value
$\overline{\langle \phi|O_t| \phi\rangle}$.

Until now, the vectors $|\phi\rangle$ 
in (\ref{1}) are in general not 
normalized. But, as seen below (\ref{12}),
the vast majority among them is 
almost of unit length.
Hence, if we replace every given
$|\phi\rangle$ in (\ref{1}) by 
its strictly normalized counterpart
\begin{eqnarray}
|\psi \rangle := 
\langle\phi|\phi\rangle^{-1/2} \, |\phi\rangle 
\ ,
\label{15}
\end{eqnarray}
then the ``new'' expectation values
$\langle \psi (t)|O| \psi (t)\rangle$
will mostly remain very close to
the ``old'' ones, i.e., to
$\langle \phi(t)|O| \phi(t)\rangle$
(since $\langle \psi (t)|O| \psi (t)\rangle=
\langle \phi(t)|O| \phi(t)\rangle/\langle\phi|\phi\rangle$
and $\langle\phi|\phi\rangle\simeq 1$).
The latter, in turn, are close to
$\overline{\langle \phi|O_t| \phi\rangle}$
for most $|\phi\rangle$'s (see below (\ref{14})).
According to (\ref{4}),
$\overline{\langle \phi|O_t| \phi\rangle}$
can be rewritten as $\tr\{\rho\, O_t\}$
and hence as
$\tr\{\rho(t)\, O\}$, 
where 
\begin{eqnarray}
\rho(t):=U_t\rho\, U_t^\dagger
\label{16}
\end{eqnarray}
is the time-evolved density operator.
Altogether, we thus can conclude that
\begin{eqnarray}
\langle \psi (t)|O| \psi (t)\rangle
\simeq
\tr\{\rho(t)\, O\}
\label{17}
\end{eqnarray}
is fulfilled in very good approximation
for most normalized vectors 
$|\psi \rangle$ in (\ref{15}).
A fully quantitative version of the latter 
approximative statement in terms of the small 
parameter $\pu$ from (\ref{10}) can be readily 
obtained by a similar line of reasoning 
as in Sec. III.C of Ref. \cite{rei18}.

Note that for any given $t$ there may still be
a small set of ``untypical'' initial states 
$|\psi \rangle$, for which  (\ref{17}) is a 
bad approximation.
Moreover, this set will usually
be different for different time points $t$.
Likewise, for any given observable $O$, the
small set of untypical states $|\psi \rangle$ 
will in general be different.
Apart from that, the main implication of
(\ref{17}) is that initial states 
$|\psi \rangle$, randomly sampled according 
to (\ref{1}) and (\ref{15}), are very likely 
to exhibit very similar expectation values 
at the initial time $t=0$ and also
at any later time, a property which 
was named {\em dynamical typicality}
in Ref. \cite{bar09}.
Formally, this typical time-evolution 
is given by the right hand side 
of (\ref{17}), but its explicit, 
quantitative evaluation is usually 
very difficult and not further 
pursued here.

\section{Basic properties of $R$ and $\rho$}
\label{s5}
So far, the only restriction regarding 
the choice of $R$ in  (\ref{1}) is that 
$\rho$ from (\ref{2}) must satisfy (\ref{3})
and (\ref{10}). 
Two further basic properties of $R$ are:

(i) Since the ensemble of vectors $|\phi\rangle$
in (\ref{1}) is invariant under arbitrary 
unitary transformations $U:\hr\to\hr$
of the basis $|\chi_n\rangle$ (see below (\ref{3})),
the same applies to the ensemble of 
vectors $|\psi \rangle$ in (\ref{15}).
In other words, $R$ and $R':=RU$ 
lead to identical dynamical 
typicality properties for arbitrary $U$.
(ii) According to textbook linear algebra, any 
linear operator $R:\hr\to\hr$ can be written
in the form $R=GU$ with a positive
semidefinite Hermitian 
operator $G$ and a unitary 
operator $U$
(so-called left polar decomposition).

Due to (i) and (ii), we can and will restrict 
ourselves Hermitian operators $R$, i.e.,
\begin{eqnarray}
R=R^\dagger 
\label{28}
\end{eqnarray}
without any loss of generality. 
Moreover, we can assume that $R$ is 
a {\em strictly} positive 
Hermitian  operator whenever this is 
convenient (which is often but 
not always the case).
The reason is as follows:
The possibility to choose $R$ 
positive {\em semidefinite} follows from 
the polar decomposition mentioned 
in (ii) above.
Furthermore,
by choosing for $|\chi_n\rangle$
in (\ref{1}) the eigenvectors of 
$R$ one sees that 
vanishing eigenvalues of $R$ are 
trivial and can be excluded right 
from the beginning by properly 
restricting the considered Hilbert 
space $\hr$.

Denoting the eigenvalues and eigenvectors
of the density operator $\rho$ by $p_n$ 
and $|n\rangle$, i.e.,
\begin{eqnarray}
\rho=\sum_{n=1}^N p_n\, |n\rangle\langle n|
\ ,
\label{18a}
\end{eqnarray}
it follows that $p_n\geq 0$,
$\sum_{n=1}^N p_n=1$ (see also (\ref{3})),
and
\begin{eqnarray}
\pu =\sum_{n=1}^N p_n^2 \ .
\label{18}
\end{eqnarray}
Moreover, the relation between $R$ and
$\rho$ may be considered as one-to-one 
in the following sense:
For any given $\rho$,
there exist exactly one Hermitian, 
semipositive $R$, which reproduces 
$\rho$ via (\ref{2}), namely
\begin{eqnarray}
R=\rho^{1/2}=\sum_{n=1}^N\sqrt{p_n}\,|n\rangle\langle n | 
\ .
\label{29}
\end{eqnarray}
Moreover, for every other $R$, 
which yields the same $\rho$ in (\ref{2}), 
the resulting random vector
ensemble in (\ref{1}) is indistinguishable 
from the ensemble generated by (\ref{29}).

In the last identity in (\ref{29}), we adopted the usual 
definition that for 
an
arbitrary function $f(x):\RR\to\RR$ 
(here $f(x)=x^{1/2}$)
and any Hermitian operator $C:\hr\to\hr$ 
with eigenvalues $c_n$ 
and eigenvectors $|\varphi_n\rangle$
(here $p_n$ and $|n\rangle$),
the operator $f(C):\hr\to\hr$ 
is defined as 
\begin{equation}
f(C):=\sum_{n=1}^N f(c_n)\, 
|\varphi_n\rangle\langle \varphi_n| 
\ .
\label{30}
\end{equation}

Focusing on strictly positive $R$'s (see below (\ref{28})),
it follows from (\ref{29}) that
$p_n>0$ for all $n$.
As a consequence, also $\rho$ 
in (\ref{18a}) is a positive
operator with a well-defined inverse, namely
\begin{eqnarray}
\rho^{-1}=\sum_{n=1}^N p_n^{-1}\, |n\rangle\langle n|
\ .
\label{30a}
\end{eqnarray}

\section{Preconditions for dynamical typicality}
\label{s3}
Until now, only the pure states 
$|\phi \rangle$ from 
(\ref{1}) and their normalized counterparts
$|\psi\rangle$ from  (\ref{15}) were of 
immediate physical
relevance, while the density operator 
$\rho$ from (\ref{2})
was mainly an auxiliary formal tool.
Additional insight into the physical 
meaning of $\rho$ can be gained 
via the precondition (\ref{10}) for 
dynamical typicality:

As observed in \cite{lin09},
the so-called effective dimension 
$\deff:=1/\pu$ tells us, how many pure 
states contribute appreciably 
to the mixture $\rho$.
Indeed, 
one readily finds -- similarly 
as in (\ref{4}) -- that
$\overline{|\phi\rangle\!\langle\phi |}=\rho$,
i.e., if this average were dominated by
just a few pure states, then the
purity $\pu$ of the mixture $\rho$
could not be very small.
More specifically, if $p_n=1/M$ for $M$ 
of the indices $n$ in (\ref{18a}),
and $p_n=0$ for all the others,
then $\deff=M$, and the $|\phi\rangle$ in 
(\ref{1}) arise by sampling vectors 
in an unbiased way within an 
$M$-dimensional subspace of $\hr$.
In other words,
$\deff$ quantifies the
``diversity'' of random vectors 
$|\phi\rangle$ contributing to $\rho$,
and (\ref{10}) ensures that
the ensemble of random vectors 
in (\ref{1}) is still 
``reasonably large''.

So far, (\ref{10}) 
represents a {\em sufficient} condition 
for the dynamical typicality properties 
established in Sec. \ref{s2}.
Next we turn to the question in 
how far this condition is 
{\em necessary} for dynamical typicality.
Our first remark is that for some
operators $B$ (e.g. the identity), 
dynamical typicality is trivially fulfilled,
independently of whether $\pu $
satisfies (\ref{10}) or not.
Hence our goal will be to show that if 
$\pu $ is not small then there exist at least
one $B$ for which dynamical typicality 
breaks down.

Taking into account (\ref{18a}) and 
(\ref{18}), one readily infers that
\begin{eqnarray}
& & p^2_{max} \leq  \pu \leq p_{max}
\ ,
\label{19}
\\
& & p_{max}  :=  \max_n p_n 
\ .
\label{20}
\end{eqnarray}
In other words $p_{max}$ is the largest
eigenvalue of $\rho$.
From (\ref{19}) we can conclude 
that $\pu $ is small if 
and only if $p_{max}$ is small,
i.e., condition (\ref{10}) 
is equivalent to
\begin{eqnarray}
p_{max}\ll 1 
\ .
\label{21}
\end{eqnarray}

Next we assume that (\ref{10}) and 
hence (\ref{21}) is {\em not} satisfied,
i.e. there is at least one $p_n$,
say $p_1$, which is not small
(compared to unity).
Choosing $|\chi_n\rangle:=|n\rangle$ 
in (\ref{1}),
$B:=|1\rangle\langle 1|$,
and observing (\ref{15}), (\ref{29}),
it follows that
\begin{eqnarray}
\langle\psi |B|\psi \rangle
& = &
\frac{p_1|c_1|^2}{\sum_{n=1}^N p_n |c_n|^2}
=
\frac{1}{1+z}
\ ,
\label{23}
\\
z & := & \frac{\sum_{n=2}^N p_n |c_n|^2}{p_1|c_1|^2}
\ .
\label{24}
\end{eqnarray}

As said below (\ref{1}), the $c_n$ are 
independent, normally distributed 
(complex) random numbers.
Dynamical typicality means that the
expectation values in (\ref{23}) 
must be very similar for most realizations
of those random numbers.
This is trivially the case if
$p_1=1$ (and hence $p_n=0$
for all $n\geq 2$).
One readily sees that this is not
a special feature of our specific 
choice of $B$ in (\ref{23})
but rather applies for
arbitrary $B$.
Similar conclusions remain true
as long as $p_1$ is very close 
to unity.
In other words, yet another sufficient 
condition (besides (\ref{21}))
for dynamical typicality is
\begin{eqnarray}
1-p_{max}\ll 1 
\ .
\label{25}
\end{eqnarray}
Once again, one can infer from (\ref{19})
that (\ref{25}) is equivalent to
\begin{eqnarray}
1-\pu \ll 1 
\ .
\label{26}
\end{eqnarray}

If $p_1$ is neither 
close to zero nor close to unity,
then one can infer that $z$
in (\ref{24}) 
is a random number, which is
typically neither very small nor 
very large and exhibits non-negligible
random fluctuations.
Hence, also the expectation value in
(\ref{23}) cannot be almost equal
for most realizations of the $c_n$'s,
i.e., dynamical typicality is violated.

In conclusion, any of the four conditions 
(\ref{10}), (\ref{21}), (\ref{25}), 
and (\ref{26})
is sufficient for dynamical typicality.
In turn, if (\ref{10}) or (\ref{21})
and simultaneously (\ref{25}) or (\ref{26})
are violated, then dynamical typicality
breaks down.
However, cases to which (\ref{25}) or 
(\ref{26}) apply are of very limited
physical interest and henceforth 
ignored.
In conclusion, either of the two
conditions (\ref{10}) or (\ref{21})
can thus be considered as 
necessary and sufficient for
dynamical typicality.

\section{Physical observables and measurement outcomes}
\label{s4}
As already pointed out at the beginning of Sec. \ref{s2},
the considered Hilbert space $\hr$ may be either
of infinite or of large but finite dimensionality.
In the first part of this section, 
we mainly have in mind the latter case.
The last part applies to both cases.

Even in case of a finite-dimensional $\hr$, 
the original Hilbert space $\hr\oo$ 
of the actual system of interest is {\em a priori}
often of infinite dimension and the
actual observables $O\oo$ are 
Hermitian operators on $\hr\oo$.
The same applies to the Hamiltonian 
$H\oo(t)$, which in general may be 
explicitly time-dependent 
(see above (\ref{13})).

In contrast to this original setup, 
a Hilbert space $\hr$ with finite dimensionality 
$N$ as introduced above (\ref{1}) usually
represents an energy shell, i.e., it is 
spanned by $N$ eigenvectors of $H\oo(0)$, 
whose eigenvalues are contained 
in an energy interval, which is microscopically
large (thus $N$ is large) but 
macroscopically small (well 
defined system energy).
Denoting by $P\oo$ the projector from
$\hr\oo$ onto $\hr$, all normalized
pure states $|\psi\rangle$ in (\ref{1}),
(\ref{15}) thus satisfy 
$P\oo|\psi\rangle=|\psi\rangle$
and 
$\langle\psi|O\oo|\psi\rangle=\langle\psi|O|\psi\rangle$
with $O:=P\oo O\oo P\oo$.
The same properties apply to $H\oo(0)$
(instead of $O'$).
For time independent Hamiltonians,
i.e. $H\oo(t)=H\oo(0)$ for all $t$,
it is therefore sufficient to focus 
on the Hilbert space $\hr$ and on 
observables $O\,:\hr\to\hr$.
This fact has been tacitly 
anticipated in Sec. \ref{s2}.
Some additional formal subtleties
in the case of explicitly time-dependent 
Hamiltonians are worked out 
in Appendix A.

It should be emphasized that
the eigenvalues and eigenvectors 
of the original (physical) observables
$O\oo$ may in general be rather
different from those of the 
auxiliary Hermitian operators $O$.
In particular, if $Q\oo:\hr\oo\to\hr\oo$
is a projector, then 
\begin{eqnarray}
Q:=P\oo Q\oo P\oo
\label{27}
\end{eqnarray}
is in general 
no longer a projector.

Note that $[O',H'(0)]=0$ (commutator) implies
$[O,H(0)]=0$ (conserved quantity),
but for the rest, commutators 
of primed and unprimed
operators are in general unrelated.

Adopting the simplest and most common 
interpretation of quantum mechanics,
the outcome of a 
measurement process will be one of the
eigenvalues of the observable $O\oo$ 
under consideration.
Both the set of possible measurement 
outcomes and their probabilities 
are usually different for $O\oo$ 
and $O$.
Only the mean values
happen to coincide for
all $|\psi\rangle\in\hr$,
while for instance the 
second moment
$\langle\psi|(O\oo)^2|\psi\rangle$
is usually (unless $O\oo=O$)
larger than
$\langle\psi|O^2|\psi\rangle$,
and likewise for the variances
(this follows from
the Cauchy-Schwarz inequality).
Nevertheless, higher moments
and thus the entire statistics 
of the measurement outcomes 
of the physical observable 
$O\oo$ are
still accessible solely by means 
of suitable auxiliary operators:
Namely, the $k$-th moment 
$\langle\psi |(O\oo)^k|\psi\rangle$
can be recovered as
$\langle \psi |\tilde O |\psi\rangle$
where the auxiliary operator
$\tilde O : \hr\to\hr$ is 
given by $P\oo (O\oo)^k P\oo$
(which usually differs 
from $O^k$).

After these preparatory considerations, 
we return to the general case of either
finite- or infinite-dimensional Hilbert 
spaces $\hr$
(in particular, $\hr$ may or may not 
agree with $\hr\oo$, and likewise for
$O$ and $H(t)$).
Our first observation is
that the dynamical typicality results 
below (\ref{17}) can also be 
applied simultaneously to 
several (possibly auxiliary)  
observables, provided their 
number is not too large.
Hence, analogous typicality properties
remain true in good approximation 
even for the entire measurement 
outcome statistics of a given 
observable $O$.
Moreover, if $O$ actually represents an 
auxiliary observable, then the same 
conclusion remains true even for the 
underlying physical observable $O\oo$.

\section{Equilibrium initial conditions}
\label{s6}
Our next goal is to identify particularly 
interesting Hermitian operators $R$ in 
(\ref{1}), (\ref{15}) which fulfill  (\ref{3}) and 
either (\ref{10}) or (\ref{21}).

The simplest case arises for time independent
Hamiltonians $H$ and operators $R$
of the form $R=f(H)$.
Hence, $R$ commutes with the time-evolution
operator $U_t=\exp\{-iHt/\hbar\}$ (see also above
(\ref{13})) and thus all statistical properties of
the random vector ensemble in (\ref{1})
are time independent.
Accordingly, the random vectors from
(\ref{1}), (\ref{15}) are called equilibrium 
initial conditions.

As usual in statistical
physics, it is not the single 
realizations $|\psi\rangle$ in 
(\ref{1}), (\ref{15}),
but rather it is the statistical 
ensemble $\rho$ in (\ref{2})
(see also (\ref{16}))
which exhibits equilibrium
(steady state) properties.
If $\rho$ satisfies (\ref{10})
or (\ref{21}),
it follows that the expectation 
value in (\ref{17})
is well approximated by the
time independent
equilibrium value $\tr\{\rho O\}$
for most $|\psi (t)\rangle$'s 
at any given time point $t\geq 0$.
Essentially, this amounts to
the original, ``non-dynamical'' or 
``canonical'' typicality results from 
\cite{llo88,gol06,pop06,sug07,rei07,gem09,sug12,tas16}.

As a first example we consider the
case when the Hilbert space $\hr$
amounts to an energy shell and,
in particular, is of large but finite 
dimensionality $N$, see also
above Eq. (\ref{1}) and Sec. \ref{s4}.
Choosing $R=\id/\sqrt{N}$ in (\ref{29})
yields for $\rho$ in (\ref{2}) 
the microcanonical ensemble
\begin{eqnarray}
\rhomic:=\id/N \ ,
\label{31}
\end{eqnarray}
satisfying (\ref{3}) and implying $\pu =1/N$.
Since we assumed $N\gg 1$ it follows that
(\ref{21}) is satisfied. 
Therefore, $\langle\psi(t)|O|\psi(t)\rangle$
will be very close to the thermal equilibrium value
$\tr\{\rhomic O\}$ for most $|\psi(t)\rangle$'s,
where $O$ is an arbitrary but fixed observable
and $t$ an arbitrary but fixed 
time point (also $t=0$ is admitted).
As noted in \cite{mal14,rei15}, the essential point
is that all possible orientations of
the orthonormal basis $|n\rangle$ in (\ref{1})
relative to the eigenbasis of $O$ are 
sampled with equal probabilities
(formally by means of Haar distributed
unitary basis transformations).
Therefore, instead of considering $|\psi(t)\rangle$
as random and $O$ as arbitrary but
fixed, one could equally well consider
$|\psi(t)\rangle$ as arbitrary but fixed and
the eigenbasis of $O$ as random 
(the eigenvalues of $O$ and the time 
point $t$ are also kept fixed).
The main conclusion is that for any given
$|\psi(t)\rangle$, the vast majority of
all observables $O$ yield expectation 
values $\langle\psi(t)|O|\psi(t)\rangle$
very close to the equilibrium value
$\tr\{\rhomic O\}$.

As a second example, we consider 
the case that $\hr$ is the full Hilbert
space of the system of interest and, 
in particular, may be infinite-dimensional,
see also above Eq. (\ref{1}) and Sec. \ref{s4}.
Choosing $R\propto\exp\{-\beta H/2\}$ 
in (\ref{29}) yields via (\ref{2}) the 
canonical ensemble
\begin{eqnarray}
\rho_{can} & := & Z^{-1}e^{-\beta H}
\ ,
\label{32}
\\
Z & := & \tr\{e^{-\beta H}\}
\ ,
\label{33}
\end{eqnarray}
where the exponential functions of $H$ 
are understood in the sense of (\ref{30}).
Furthermore, condition (\ref{21})
will be satisfied if $H$ exhibits many 
eigenvalues $E_n$, whose distance 
to the ground state energy is smaller
than the thermal energy $k_BT:=1/\beta$,
since many of the eigenvalues 
$p_n=Z^{-1} \exp\{-\beta E_n\}$ of $\rho$
will then be comparable to $p_{max}$ 
and hence $p_{max}$ must be small.
Given that generic many-body Hamiltonians
exhibit an extremely dense spectrum
even at rather low energies, condition
(\ref{21}) will thus be fulfilled for virtually
any experimentally realistic temperature
$T$.

In conclusion,
very accurate approximations for any
expectation value in the canonical 
ensemble can be obtained with extremely
high probability by randomly sampling 
a single pure state $|\psi\rangle$
according to (\ref{1}), (\ref{15})
and $R$ as specified above.
Moreover, we recall that in (\ref{1})
any basis $|\chi_n\rangle$ of the
Hilbert space under consideration 
will do.

For practical purposes, it is often convenient to
break up the above scheme into two steps.
In a first step, instead of working
with $R=\gamma \exp\{-\beta H/2\}$, 
where $\gamma$ is fixed via (\ref{2}) 
and (\ref{3}), one rather chooses 
$\gamma=1$.
The resulting operator $R=\exp\{-\beta H/2\}$
can be numerically evaluated very
efficiently for instance
by means of imaginary time 
propagator techniques
\cite{bar17}.
Obviously, high temperatures (small $\beta$) 
are expected to be particularly easily 
tractable, however, apart from the above excluded,
exceedingly low temperatures, 
all temperature regimes should be feasible.
Moreover, the values of
$\langle\phi|\phi\rangle$
for the resulting random vectors 
$|\phi\rangle$ in (\ref{1})
will no longer be with very high
probability close to unity (see below 
(\ref{12})), but rather close
the value $Z$  from (\ref{33}).
Since $Z$ is the canonical partition function,
one thus also obtains the free energy
and all the hence deriving further 
thermodynamic quantities in the usual way.
In a second step, the canonical expectation 
value of an arbitrary observable $O$ can be
obtained from the same random vector
$|\phi\rangle$ by evaluating
$\langle\phi|O|\phi\rangle/\langle\phi|\phi\rangle$.
In conclusion, a single typical pure state
$|\phi\rangle$
is able to imitate all physically relevant 
properties  of a full fledged canonical 
equilibrium ensemble\,!

All these quite remarkable observations
have been analytically elaborated 
and numerically exemplified in a 
particularly clear and comprehensive 
work by Sugiura and Shimizu in 
Ref. \cite{sug13}.
However, many of the key ideas have been
theoretically recognized and numerically 
employed in a considerable number
of much earlier works, most
notably in Ref. \cite{ham00},
but also in Refs. \cite{r2}.
Analogous concepts of how to imitate
microcanonical and grand canonical
ensembles by randomly sampling a 
single pure state have been worked 
out in Refs. \cite{sug12,hyu14}.

{\em If the Hamiltonian is explicitly 
time-dependent, already such simple choices
as $R=f(H(0))$ may lead to quite interesting
consequences and applications\,!}
The main caveat is the observation
below (\ref{31}) that most observables
may yield expectation values close to the
equilibrium value for any given 
$|\psi(t)\rangle$ \cite{mal14,rei15}.
Therefore, it may be necessary to
carefully select a suitable observable 
in order to ``see'' any non-trivial
effect of the external perturbation, 
which generates the time-dependence
of the Hamiltonian.
This may be worthwhile to be
explored numerically for
specific model systems, e.g. 
exhibiting a quench dynamics or a
periodic time-dependence.

\section{Non-equilibrium initial conditions}
\label{s7}
Especially for time independent Hamiltonians,
initial states with certain non-equilibrium
expectation values are of particular interest.
Referring to 
Sec. \ref{s11}
for a more
detailed justification, we henceforth focus
on situations where the expectation values of 
one or several observables $A_k$ ($k=1,...,K$)
are known to (approximately) assume 
certain values $\ak_k$, which differ from 
the equilibrium expectation values 
discussed in the preceding subsection.

A natural way to incorporate those preset 
expectation values $\ak_k$ is by means of 
additional constraints on $\rho$ (and thus 
via (\ref{2}) or (\ref{29}) on $R$) of the form
\begin{eqnarray}
\tr\{\rho A_k\}=\ak_k\ \mbox{for}\ k=1,...,K.
\label{34}
\end{eqnarray}
Indeed, 
taking for granted (\ref{10}) and
setting $t=0$ in (\ref{17}) 
implies that $\langle \psi |A_k| \psi \rangle$
will be close to $\ak_k$ for most 
initial states $|\psi \rangle$.

Accordingly, our next goal is to 
find $\rho$'s 
which satisfy (\ref{34})
on top of (\ref{3}) and (\ref{10}).
First of all, it seems reasonable 
to assume that the non-equilibrium 
character of a macroscopic initial 
state in a real experiment (see also 
Sec. \ref{s11}) can be 
captured by a relatively small 
number $K$ 
of non-trivial (non-equilibrium)
expectation values $\ak_k$ 
appearing in (\ref{34}).
Yet, it is quite obvious that there
still may exist certain combinations 
of $A_k$'s and $\ak_k$'s for 
which (\ref{3}) and (\ref{34}) 
cannot be satisfied by {\em any} 
(pure or mixed) quantum 
state $\rho$\,:

To begin with, only $\ak_k$ values 
in between the smallest and the largest 
eigenvalues of $A_k$ are admissible,
otherwise (\ref{34}) cannot be 
satisfied by any $\rho$.

Furthermore, the operators 
$\id$ and all the $A_k$'s
in (\ref{34}) must be
linearly independent in the 
following sense:
The only real numbers 
$x_k$ which solve 
the equation
\begin{eqnarray}
x_0\,\id + \sum_{k=1}^K x_k\, A_k = 0
\label{39}
\end{eqnarray}
are $x_k=0$.
The reason is as follows:
If non-trivial solutions 
of (\ref{39}) would exist, 
then at least one $x_k$ with
$k>0$ must be non-zero,
say $x_1\not=0$.
With $b_k:=-x_k/x_1$ it follows that
$A_1=b_0\,\id +\sum_{k=2}^K b_k A_k$
and with (\ref{3}), (\ref{34}) that 
$\ak_1=b_0+\sum_{k=2}^K b_k \ak_k$.
If $\ak_1$ fulfills this
equality then the constraint
(\ref{34}) for $k=1$ is 
redundant and can be omitted
without any loss of generality.
If $\ak_1$ violates this equality,
then the problem was ill posed 
in the first place.

Finally, it should be emphasized 
that the $A_k$'s in (\ref{34}) 
may or may not commute, and that particularly 
complicated restrictions regarding physically 
reasonable combinations of the $\ak_k$'s are 
expected to arise for non-commuting $A_k$'s.
The reasons for such incompatible 
constraints may or may not be 
obvious and their systematic exploration 
goes beyond our present scope.
Moreover, while the so far discussion
mainly emphasized the mathematical
side of the problem, 
physically speaking such incompatible 
conditions would indicate either a 
meaningless (experimentally unrealistic) 
situation or that the theoretical model 
was inappropriate in the first place.
For all these reasons, 
we henceforth take for granted 
that (\ref{3}) and (\ref{34}) are 
satisfiable by at least one density 
operator $\rho$.

In many cases, already a few constraints
of the form (\ref{34}) may adequately 
capture the non-equilibrium initial 
condition at hand.
In other cases, the number $K$ of relevant 
constraints (\ref{34}) may be quite considerable,
for instance in order to describe initial states
with complex spatial inhomogeneities.
Nevertheless, a number $K$ much
smaller than $N$ will usually
be sufficient.
In particular, the number of parameters 
(independent matrix elements)
specifying $\rho$ is thus much larger than 
the number of equations (\ref{3}) and (\ref{34}).
Moreover, (\ref{3}) and (\ref{34}) give rise to
linear equations for the matrix elements of $\rho$.
It follows that -- given at least one solution 
exists -- generically there actually will be
many different solutions $\rho$
of those equations.
Likewise, if one of those $\rho$'s in addition
satisfies condition
(\ref{10}), then the same is expected to
apply also for many other $\rho$'s
(for instance those within a 
``close neighborhood'').
Moreover, if we denote by $\rho_1$ and 
$\rho_2$ two different such solutions,
then generically there will exist certain
observables $B$ for which the expectation
values $\tr\{\rho_1 B\}$ and $\tr\{\rho_2 B\}$ 
differ by a non-negligible amount.
Recalling that the Hamiltonian and hence
the unitary time propagator $U_t$ in (\ref{14})
may, in principle, still be chosen 
largely arbitrary 
(see also Appendix \ref{app1}),
the above $B$ may be reproduced 
by $O_t$ via (\ref{14}) in terms
of a quite ``harmless'' observable $O$ 
(for instance by one of the $A_k$'s),
yielding $\tr\{\rho\,B\}$ for the
expectation value at time $t$ on the 
right hand side of (\ref{17}).
An extreme example arises when both 
$\rho_1$ and $\rho_2$ satisfy
(\ref{3}), (\ref{10}),  (\ref{34}),
and, in addition, both ``secretly''
satisfy one more condition of the 
type (\ref{34}), 
say $\tr\{\rho B\}=\beta$,
but with two different values 
for $\beta$.
In this case, dynamical typicality behavior
is recovered in both cases, but the typical
expectation values are different in the two 
cases, for instance for the observable 
$B$ at time $t=0$.

We also note that whenever two different
solutions $\rho_1$ and $\rho_2$ of (\ref{34})
are at hand, then most of the random vectors
associated with $\rho_1$ via (\ref{1}), 
(\ref{15}), and (\ref{29})
must be considered as ``untypical'' samples
of the random vector ensemble associated 
with $\rho_2$.

\subsection{Unbiasedness assumption}
\label{s71}
The overall so far conclusion is: 
Given that (\ref{3}), (\ref{10}), and (\ref{34})
can be satisfied at all, then 
generically many different 
$\rho$'s will do so.
While each of them will 
exhibit  dynamical typicality, 
the predicted typical behavior
may quite notably differ between
some of them.
Hence, the question arises,
which $\rho$ is the ``right'' one,
e.g., in order to describe an ensemble of 
experiments which are repeated (as well 
as possible) at different times or in 
different labs (see Sec. \ref{s11}).

In order to uniquely determine the ``right'' 
 $\rho$ along these lines, 
additional requirements are clearly indispensable.
Here we choose them as announced in Sec. \ref{s11}
(see also above Eq. (\ref{1}), at the beginning 
of Sec. \ref{s4}, and around (\ref{31})).
Namely, we assume that the initial system energy 
is known (and reproducible) up to an uncertainty 
which is small on macroscopic but large on microscopic 
scales, i.e., the relevant Hilbert space $\hr$ is
an energy shell, spanned by a large but finite 
number $N$ of eigenvectors of the 
(time independent) Hamiltonian $H$.
Within this Hilbert space, our random ensemble
of vectors $|\psi\rangle$ is defined as usual 
via (\ref{1}) and (\ref{15}).
Now our key requirement, by which $\rho$ will then be 
uniquely determined, consist in the
following unbiasedness assumption:
{\em If two random vectors $|\psi\rangle$ exhibit equal 
expectation values $\langle\psi|A_k|\psi\rangle$ 
for all $k=1,...,K$, then they are realized with 
equal probability.}
It is our contention that the situation 
outlined in Sec. \ref{s11} is appropriately 
captured in this way:
We know the (approximate) energy 
of the initial system state $|\psi\rangle$, as 
well as the (approximate) expectation 
values of the observables $A_k$.
Apart from that, no further information
about $|\psi\rangle$ is available.
It is therefore natural to model the initial
states in accordance with the above 
unbiasedness assumption, e.g., in order
to describe many repetitions of the 
``same'' experiment.

Yet another important justification of the 
above unbiasedness assumption is that 
the resulting random vectors $|\psi\rangle$ 
will closely approximate the random 
vectors with strictly fixed initial conditions 
from Sec. \ref{s81}.

\subsection{Determination of $\rho$}
\label{s72}
Our starting point is the following 
general result, whose detailed derivation
is provided in Appendix B:
Given an arbitrary but fixed $\rho$,
we choose its eigenvectors
$|n\rangle$ (cf. (\ref{18a}))
as basis vectors 
$|\chi_n\rangle$ in (\ref{1}).
With (\ref{29}) we thus can rewrite
(\ref{15}) as
\begin{eqnarray}
|\psi\rangle & = & \sum_{n=1}^N y_n |n\rangle
\ ,
\label{n1}
\\
y_n & = & \frac{\sqrt{p_n}\, c_n}{\sqrt{\sum_{n=1}^N p_n |c_n|^2}}
\ .
\label{n2}
\end{eqnarray}
By definition, the probability distribution 
$\tilde w(|\psi\rangle)$ of the vectors $|\psi\rangle\in\hr$ 
is equal to the probability distribution 
$w({\bf y})$
of the corresponding random numbers
${\bf y}:=(y_1,....,y_N)$  in (\ref{n1}), 
whose statistical properties are
determined via (\ref{n2}) by those of the $c_n$'s 
in (\ref{1}).
Referring to Appendix B for the details, the final 
result is
\begin{eqnarray}
\tilde w(|\psi\rangle)
& = & w({\bf y}) =
\nn\, \frac{\delta(\norm{\psi}-1)}{\langle\psi|\rho^{-1}|\psi\rangle^{N}}
\ ,
\label{n3}
\\
\nn & := &\frac{(N-1)!}{2\,\pi^N\det(\rho)}
\ ,
\label{n4}
\end{eqnarray}
where $\norm{\psi}:=\langle\psi|\psi\rangle^{1/2}$ is the
norm of $|\psi\rangle$ and
$\det(\rho):=\prod_{n=1}^Np_n$ the 
determinant for $\rho$ from (\ref{18a}).
Moreover, we have taken for granted that the considered
Hilbert space $\hr$ has already been properly adapted
as detailed below (\ref{28}), so that $\rho$ is strictly
positive and its inverse in (\ref{30a}) is well defined.

We remark that the same result as in (\ref{n3}) has been
obtained (by methods different from those in Appendix B)
previously in Ref. \cite{rei19} (see Appendix therein).
Moreover, a closely related result is also contained
in Ref. \cite{gol06a} (see equation (18b) therein).

Focusing on $\rho$'s which satisfy the unbiasedness assumption
from Sec. \ref{s71}, the probability in (\ref{n3})  must be equal 
for all normalized vectors $|\psi\rangle$ with identical 
values of $a_\psi^k:=\langle\psi |A_k|\psi\rangle$.
It follows that the quantity $\langle\psi|\rho^{-1}|\psi\rangle$
on the right hand side of (\ref{n3}) must be completely
determined by the $K$ real numbers $a_\psi^k$, i.e.,
there exists a function $h:\RR^K\to\RR$
with the property that
\begin{eqnarray}
\langle\psi|\rho^{-1}|\psi\rangle 
= 
h(a_\psi^1,...,a_\psi^K)
\label{n5}
\end{eqnarray}
for all normalized vectors $|\psi\rangle$.
Under the tacit assumption that the 
function $h(x_1,...,x_K)$ is differentiable 
at $x_k=0$ ($k=1,...,K$),
we can conclude that there 
must exist real constants 
$\lambda_k:=\partial h(0,...,0)/\partial x_k$
so that
\begin{eqnarray}
\langle\psi|\rho^{-1}|\psi\rangle 
& = &
\lambda_0 + \sum_{k=1}^K \lambda_k\, 
\langle\psi |A_k|\psi\rangle
+...
\nonumber
\\
& = &
\langle\psi|\, \bigg(\lambda_0 \id + \sum_{k=1}^K \lambda_k\, 
A_k\bigg)\, |\psi\rangle
+...
\label{n6}
\end{eqnarray}
for all normalized vectors $|\psi\rangle$, 
where the dots indicate
the usual ``remainder terms'' of
higher than linear order in 
$a_\psi^k:=\langle\psi |A_k|\psi\rangle$.
It is reasonable to expect that, in general, 
these higher order contributions cannot be 
written in the form
$\langle\psi |B|\psi\rangle$ for some 
Hermitian operator $B$, unless they happen 
to be identically zero.
But since all other operators appearing in
(\ref{n6}) are Hermitian, those
higher order terms must be zero.
The remaining equation (\ref{n6})
can only be valid for all
$|\psi\rangle$ if the operators 
on both sides are identical,
implying
\begin{eqnarray}
\rho & = & \left[ \lambda_0 \id + \sum_{k=1}^K \lambda_k\, A_k\right]^{-1}
\ ,
\label{n7}
\end{eqnarray}
where the right hand side is understood in 
the sense of (\ref{30}).
An alternative, more rigorous 
but also more involved derivation of 
(\ref{n7}) is provided in Appendix C.

The remaining $K+1$ parameters 
$\lambda_0,...,\lambda_K$
in (\ref{n7}) must be chosen so that 
the $K+1$ equations (\ref{3}) and (\ref{34})
are fulfilled.
Moreover, the resulting solution 
must satisfy
\begin{eqnarray}
\lambda_0\,\id+\sum_{k=1}^K\lambda_k\, A_k > 0
\ .
\label{91}
\end{eqnarray}
since $\rho$ was assumed to be
a strictly positive operator
(see below (\ref{n4})).
Regarding the issue whether any solution
of those equations exists, similar 
considerations as below (\ref{34}) apply.
In particular, we again restrict ourselves
to situations, for which at least one 
solution exists.
But since the number of parameters is now 
equal to the number of equations, it is 
reasonable to expect that this solution 
will moreover be unique. 
In the special case $K=1$ this has been 
rigorously verified in \cite{rei18}.
Finally, one of the $K+1$ equations can 
always be solved right away so that
effectively only $K$ equations 
with $K$ unknowns remain, 
see Appendix \ref{app4}.

The above findings represent the first main 
result of our paper. In particular,
any $\rho$ which satisfies (\ref{3}) and (\ref{34})
but is {\em not} of the form (\ref{n7}) is biased:
Random vectors $|\psi\rangle$ with
identical expectation values for all $A_k$ are
{\em not} realized with equal probabilities.
If in addition (\ref{10}) is fulfilled, such a $\rho$
still exhibits dynamical typicality properties,
but they may be 
different from those which an 
unbiased ensemble would exhibit.

\subsection{Canonical density operators are biased}
\label{s73}
As stated at the beginning of Sec. \ref{s71},
from the physical viewpoint we are mainly 
interested in the ``microcanonical''
situation that the relevant Hilbert 
space $\hr$ amounts to an energy shell.
However, from the formal viewpoint, 
our above derived 
main results remain valid under 
considerably more general circumstances.
In particular, we may consider the special 
case of a time-independent system (Hamiltonian $H$)
with only one single constraint ($K=1$)
of the form $\tr\{\rho H\}=E$,
i.e., $A_1:=H$ and $\alpha_1:=E$.
In this case, one
might naively expect that the resulting
density operator $\rho$ should be
of the ``canonical'' form $Z^{-1}e^{-\beta H}$
for some suitable $\beta$ and $Z$
(provided $E$ was chosen ``reasonable'').
On the other hand, such a canonical 
form is clearly incompatible with 
(\ref{n7}).
The reason is that such a canonical 
density operator $\rho$ violates the 
unbiasedness assumption from 
Sec. \ref{s71}.
At first glance, this may appear hard to
believe, but upon closer inspection one 
realizes that one actually cannot come up 
with a good argument why a canonical 
$\rho$ should generate an unbiased 
ensemble of random vectors.
In fact, one can readily construct simple 
examples which explicitly show that such 
ensembles are biased. 
For more details on this issue 
see also \cite{fine09}.

\section{Alternative ensembles}
\label{s8}
We start by recalling the three main
properties of the normalized random 
vectors $|\psi\rangle$ in (\ref{15}) 
or (\ref{n1}) when the density operator 
is of the specific form (\ref{n7}) and
satisfies (\ref{3}), (\ref{10}), and
(\ref{34}):
(i) Dynamical typicality:
For most of the 
time-evolved states $|\psi(t)\rangle$,
the expectation values 
$\langle \phi(t)|O|\phi(t)\rangle$ 
are close to the ensemble average in 
(\ref{17}) for any given observable 
$O$ and time point $t$.
(ii) Initial disequilibrium:
The expectation values 
$\langle\psi|A_k|\psi\rangle$ 
at time $t=0$
are {\em almost} equal to the preset
ensemble averages in (\ref{34}) 
for {\em most} of the 
$|\psi\rangle$'s.
(iii) Unbiasedness:
{\em All} vectors $|\psi\rangle$ with 
{\em exactly} identical expectation 
values $\langle\psi|A_k|\psi\rangle$ 
for all $k=1,...,K$ are realized with
equal probability. 

Besides this random vector ensemble,
henceforth denoted as $E$, the following 
subsections explore two alternative  
ensembles of normalized random
vectors, which will be named $S$ and GAP, 
and which exhibit the same
general properties (i)-(iii) as $E$.
More precisely, with respect to (i) and (ii),
all their statistical properties are
practically indistinguishable from those 
of $E$, though in principle there remain
certain, extremely small deviations,
while (iii) is strictly obeyed by all 
three ensembles.

\subsection{Ensembles with strictly fixed initial conditions}
\label{s81}
We consider the ensemble $S$ 
of all normalized vectors $|\psi\rangle$, 
whose expectation values
$\langle\psi|A_k|\psi\rangle$ 
are {\em strictly} (hence the ``S'')
equal to some preset values $x_k$, i.e.,
\begin{eqnarray}
\langle\psi|A_k|\psi\rangle & = & x_k
\ \mbox{for}\ k=1,...,K.
\label{m1}
\end{eqnarray}
For the rest, all those normalized
$|\psi\rangle\in S$ are again 
realized with equal probability.
Note that {\em a priori}, the $x_k$ 
in (\ref{m1}) are independent of the
$\alpha_k$ in (\ref{34}), 
but it will obviously be of foremost
interest to compare ensembles $E$ 
and $S$ with $x_k\simeq \alpha_k$.
For $K=1$,
the ensemble $S$ and many of its principal 
properties have been previously 
explored by 
Fine \cite{fine09} and by M\"uller et al. 
\cite{mueller11}, see also \cite{rei18}.

Put differently, $S$ fulfills the
same property (iii) as before in 
conjunction with a property, which 
is stronger than (ii).
An interesting question is therefore, 
whether $S$ still exhibits dynamical
typicality (property (i)).
In the following, we will argue that
this is indeed the case.
Moreover, when $x_k\simeq \alpha_k$
for all $k$ then the corresponding
typical expectation values 
$\langle \psi(t)|O|\psi(t)\rangle$ 
for the two ensembles $S$ and $E$
will even be very 
similar to each other.
In other words, most random vectors
from one ensemble closely imitate
the behavior of most random
vectors from the other ensemble.

Conceptually, the ensemble $S$  seems in 
many respects physically more sensible 
than $E$. 
For instance, its properties are in some way 
``cleaner'' (more precisely specified)  and it
contains ``fewer'' elements $|\psi\rangle$, 
so that statements about ``most'' $|\psi\rangle$ 
are in a certain sense ``stronger''.
On the other hand, finding actual samples
$|\psi\rangle$ of $S$ in practice (e.g. by 
numerical means), seems considerably 
more demanding than sampling random
vectors $|\psi\rangle$ from $E$.
Therefore, it is very useful to know that
a typical $|\psi\rangle\in E$ imitates 
very well the properties of a typical 
$|\psi\rangle\in S$.

Similarly as in (\ref{13}) and (\ref{14}), 
taking into account the time-evolution
of $|\psi\rangle$ can be avoided by
exploring the typicality properties
of the expectation values 
$\langle\psi|B|\psi\rangle$
for an arbitrary but fixed 
observable $B$.
Furthermore, the joint probability
density
$P_E(b,{\bf x})$ that a random vector
$|\psi\rangle\in E$ 
simultaneously satisfies
$\langle\psi|B|\psi\rangle=b$ and 
$\langle\psi|A_k|\psi\rangle=x_k$
for arbitrary but fixed values of
$b$ and ${\bf x}:=(x_1,...,x_K)$
can be written by means of the probability
distribution from (\ref{n3}) as
\begin{eqnarray}
P_E(b,{\bf x}) \!\! & = & \!\! \int d{\bf y}\, w({\bf y})\ 
\delta(b_\psi-b)
\prod_{k=1}^K \delta(a_\psi^k - x_k)
\ ,
\label{m2}
\\
b_\psi & := & \langle\psi|B|\psi\rangle
\ ,
\label{m2a}
\\
a_\psi^k & := & \langle\psi|A_k|\psi\rangle
\ .
\label{m2b}
\end{eqnarray}
Exploiting in (\ref{n3}) the special form (\ref{n7}) of
$\rho$ for the ensemble $E$ under consideration, 
we can rewrite (\ref{m2}) as
\begin{eqnarray}
\!\! P_E(b,{\bf x}) \!\!\! & = & \!\!\! g({\bf x}) \!\!
\int\!\! d{\bf y}\, \delta(\norm{\psi}\!-\!1)\,\delta(b_\psi-b)
\!\prod_{k=1}^K\! \delta(a_\psi^k-x_k)
\ \ \ \ \ 
\label{m3}
\\
\!\!g({\bf x}) \!\! & := & \!\! \nn
\left(\lambda_0+\mbox{$\sum_{k=1}^K$}\lambda_k\, x_k\right)^{-N}
\ ,
\label{m3b}
\end{eqnarray}
where the normalization constant $\nn$ 
is defined in (\ref{n4}).
As usual, the (marginal) probability distribution
for the expectation values of $B$ alone 
follows by integrating out the variables ${\bf x}$, i.e.,
\begin{eqnarray}
P_E(b):=\int d{\bf x}\, P_E(b,{\bf x})
\ .
\label{m4}
\end{eqnarray}
Similarly, the (joint) probability distribution for the 
expectation values of the $A_k$'s alone is
\begin{eqnarray}
\tilde P_E({\bf x}):=\int db\, P_E(b,{\bf x})
\ .
\label{m5}
\end{eqnarray}
The only purpose of the tilde symbol is to better
distinguish the two quantities on the left hand side
of (\ref{m4}) and of (\ref{m5}).
Finally, the conditional probability $P_E(b|{\bf x})$,
i.e. the probability that $\langle\psi|B|\psi\rangle$
assumes the value $b$, given that 
$\langle\psi|A_k|\psi\rangle=x_k$
for all $k$, is defined as
\begin{eqnarray}
P_E(b|{\bf x}) := \frac{P_E(b,{\bf x})}{\tilde P_E({\bf x})}
\ .
\label{m6}
\end{eqnarray}
Introducing (\ref{m3}) into (\ref{m5}) and (\ref{m6}) yields
\begin{eqnarray}
P_E(b|{\bf x})=
\frac{
\int\! d{\bf y}\, \delta(\norm{\psi}-1)\ \delta(b_\psi-b)
\prod_{k=1}^K \delta(a_\psi^k-x_k)
}
{
\int d{\bf y}\, \delta(\norm{\psi}-1)\,
\,\prod_{k=1}^K \delta(a_\psi^k-x_k)
}
\ .\ \
\label{m7}
\end{eqnarray}
One readily sees that this is exactly the probability
distribution for the expectation values of $B$ when the
$|\psi\rangle$'s are randomly sampled from the
ensemble $S$ with constraints (\ref{m1}),
symbolically indicated as
\begin{eqnarray}
P_S(b|{\bf x}) = P_E(b|{\bf x}) 
\ .
\label{m7a}
\end{eqnarray}

Throughout this section, the $\alpha_k$'s 
in (\ref{34}) are tacitly considered as 
arbitrary but fixed.
However, if we would consider them as variable, 
then $P_E(b,{\bf x})$ would depend on the
$\alpha_k$'s via $g({\bf x})$ in (\ref{m3}).
As a consequence, also $P_E(b)$ in (\ref{m4})
and $\tilde P_E({\bf x})$ in (\ref{m5})
are generically expected  to depend on 
the $\alpha_k$'s.
On the other hand, $P_E(b|{\bf x})$ in 
(\ref{m6}) quite remarkably is independent
of the $\alpha_k$'s, as can be inferred
from (\ref{m7}).
Due to (\ref{m7a}), the same property
applies to $P_S(b|{\bf x})$.

Combining (\ref{m4}), (\ref{m6}), and (\ref{m7a}) 
yields
\begin{eqnarray}
P_E(b):=\int d{\bf x}\, \tilde P_E({\bf x})\, P_S(b|{\bf x})
\ .
\label{m8}
\end{eqnarray}
Taking for granted that the density operator
$\rho$, corresponding to the ensemble $E$ 
at hand, satisfies (\ref{10}), we know that 
dynamical typicality applies.
In particular, the distribution of the expectation 
values of $B$ on the left hand side in (\ref{m8})
must be sharply peaked about the
mean value $\tr\{\rho B\}$.
(Closer inspection reveals that the distribution
is actually very close to a Gaussian if $K=1$ 
\cite{rei19}, and it is reasonable to expect
that the same remains true for $K>1$.)
Likewise, the joint distribution $\tilde P_E({\bf x})$ 
of the expectation values of the $A_k$'s 
must be very narrowly peaked about the corresponding
mean values $\tr\{\rho A_k\}$, which
in turn are equal to the $\alpha_k$'s
according to (\ref{34}).
Is it possible that the convolution
of this narrow distribution $\tilde P_E({\bf x})$
with $P_S(b|{\bf x})$ on the right hand side
of (\ref{m8}) can lead to a very narrow 
distribution $P_E(b)$ on the left hand side,
without $P_S(b|{\bf x})$ being a very sharply 
peaked function of $b$ for any given 
vector ${\bf x}$\,? 
In principle, this is clearly 
possible as long as the ${\bf x}$ values 
with non-narrow functions $P_S(b|{\bf x})$  
only contribute with small weights 
$\tilde P_E({\bf x})$ in (\ref{m8}).
In practice, it seems reasonable to 
expect that the main features of the function 
(\ref{m7}) of $b$ (e.g., its mean and variance)
should not abruptly change upon small
variations of the parameters ${\bf x}$.
(Again, for $K=1$ this can even be 
rigorously verified \cite{rei18}).
Taking this assumption for granted,
one can approximate
$P_S(b|{\bf x})$ 
by $P_S(b|\boldsymbol{\alpha})$
for all ${\bf x}$ with non-negligible
weights $\tilde P_E({\bf x})$ in (\ref{m8}),
where $\boldsymbol{\alpha}:=
(\alpha_1,...,\alpha_K)$, i.e.,
\begin{eqnarray}
P_S(b|{\bf x})\simeq P_S(b|\boldsymbol{\alpha})
\ \ \mbox {if}\ \ 
{\bf x}\simeq \boldsymbol{\alpha}
\ .
\label{m9}
\end{eqnarray}
We thus can infer from (\ref{m8}) the approximation
\begin{eqnarray}
P_E(b)\simeq P_S(b|\boldsymbol{\alpha})
\ .
\label{m10}
\end{eqnarray}
These approximations (\ref{m9}) and 
(\ref{m10}) are tantamount to the
announced result that the two random 
vector ensembles $E$ and $S$ with
$x_k\simeq \alpha_k$
exhibit very similar typicality 
properties.

\subsection{The Gaussian adjusted projected ensemble}
\label{s82}
Our starting point is an arbitrary but fixed 
density operator $\rho$ in (\ref{18a}).
The corresponding, so-called Gaussian adjusted 
(GA) ensemble \cite{gol06a} consist of random 
vectors 
\begin{eqnarray}
|\varphi\rangle & = & \sum_{n=1}^N v_n\, |n\rangle
\ ,
\label{m11}
\end{eqnarray}
where the random variables ${\bf v}:=(v_1,...,v_N)\in \CC^N$
may equivalently be considered as $2N$-dimensional real 
vectors (see also Appendix \ref{app2}), and whose probability
distribution is given by
\begin{eqnarray}
w_{GA}({\bf{v}}) & := & 
\frac{1}{\pi^N\det(\rho)}
\  \norm{{\bf v}}^2\, 
\exp\left\{-\sum_{n=1}^N \frac{|v_n|^2}{p_n}\right\}
\ . \ \ \ 
\label{m12}
\end{eqnarray}
The word ``Gaussian'' refers to the 
exponential term in (\ref{m12}), 
while the word ``adjusted'' refers 
to the factor $\norm{{\bf v}}^2$, 
whose purpose will become clear later.
Similarly as in (\ref{1}), the norm of the vector 
in (\ref{m11}) is itself a random 
variable.

The so-called Gaussian adjusted projected (GAP) 
ensemble amounts 
-- similarly as in (\ref{15}) --
to the normalized counterparts of (\ref{m11}),
hence the name ``projected'' \cite{gol06a,rei08}.
In other words the GAP ensemble
consists of normalized vectors
$|\psi\rangle$ of the form (\ref{n1}), where,
instead of (\ref{n3}), the probability distribution
of the random variables ${\bf y}:=(y_1,...,y_N)$
now derives from that of the ${\bf v}$'s in (\ref{m12})
according to (see also Appendix \ref{app2})
\begin{eqnarray}
w_{G}({\bf y}) & := & \int d{\bf v}\, w_{GA}({\bf v})\ 
\delta({\bf y}-{\bf v}/\norm{\bf v})
\ .
\label{m13}
\end{eqnarray}
Similarly as in (\ref{n3}), the probability distribution
$\tilde w_{G}(|\psi\rangle)$ of the vectors 
$|\psi\rangle$ is identified with $w_{G}({\bf y})$,
and similarly as in the Appendix \ref{app2},
the integral in (\ref{m13}) can be evaluated
to yield
\begin{eqnarray}
\tilde w_{G}(|\psi\rangle)
& = &
w_{G}({\bf y}) = 
\nn_{G} 
\ 
\frac{\delta(\norm{\psi}-1)}
{\langle\psi |\rho^{-1} |\psi \rangle^{N+1}}
\ ,
\label{m14}
\\
\nn_{G} & := &
\frac{N!}{2\,\pi^N\det(\rho)}
\ .
\label{m15}
\end{eqnarray}

Denoting the average over the above specified GAP
ensemble by an overbar,
it has been shown in Refs. \cite{gol06a,rei08} that
\begin{eqnarray}
\bar B & := & \overline{\langle\psi|B|\psi\rangle}
= \tr\{\rho  B\}
\ ,
\label{m16}
\\
\sigma_B^2 & := &
\overline{
\left(\langle\psi|B|\psi\rangle
-\overline{\langle\psi|B|\psi\rangle}\right)^2
}
\nonumber
\\
& \leq &
\db^2\, \pu\left(1+\ord(\sqrt{\pu}\,)\right)
\ ,
\label{m17}
\end{eqnarray}
where $B:\hr\to\hr$ may be any Hermitian
operator, and
where $\db$ denotes the range of $B$,
i.e., the difference between its largest 
and smallest eigenvalues.

These results are remarkably similar to
those in (\ref{4}) and (\ref{5}).
The main difference is that the 
norm of the vectors $|\phi\rangle$
in (\ref{4}) and (\ref{5}) is a random
variable, while the vectors
$|\psi\rangle$ in (\ref{m16}) 
and (\ref{m17}) are strictly 
normalized.
This is the main purpose of the ``adjustment''
factor ${\norm{\bf v}}^2$, as announced 
below (\ref{m12}).
Note also that,
as discussed around (\ref{17}), the 
normalized counterparts of the 
$|\phi\rangle$'s from (\ref{15}) 
satisfy a relations of the form  (\ref{m16}) 
only approximately, and under the additional 
proviso that $\rho$ satisfies the extra
condition (\ref{10}).
In contrast, (\ref{m16}) is an exact identity
for arbitrary density operators $\rho$
and arbitrary $N$.
Likewise, the term $\ord(\sqrt{\pu}\,)$
on the right hand side of (\ref{m17})
can be replaced by a rigorous but somewhat 
lengthy upper bound, see equation (75) 
in \cite{rei08}.

Under our usual premises that $\rho$ satisfies
(\ref{3}), (\ref{10}), and (\ref{34}), the
random vector ensemble 
$E$
from (\ref{1}), (\ref{15})
[or from (\ref{n1}), (\ref{n2})]
is thus practically indistinguishable 
from the GAP ensemble.
In particular, if $\rho$ is of the
form (\ref{n7}), then both ensembles
exhibit the properties (i)-(iii) 
from the beginning of this Section. 
Moreover the close similarity 
is directly indicated by the 
explicit expressions for the probability 
distributions corresponding to $E$,
as given in (\ref{n3}), and corresponding 
to GAP as given in (\ref{m14}). 
Obviously the only relevant difference 
is the exponent of the denominator, which 
is $N$ for $E$ and $N+1$ for GAP. 
For sufficiently large $N$ this difference 
is relatively minor.
\\[0.5cm]

\subsection{Discussion}
\label{s83}
In comparison with the ensemble $E$ 
(defined at the beginning 
of this section), 
a key difference of the $S$ and GAP 
ensembles is 
that  when writing the corresponding
random vectors in the form (\ref{n1})
then -- unlike in (\ref{n2}) -- 
the coefficients $y_n$ can no 
longer be conveniently realized 
by straightforward sampling of  independent 
random variables.
In particular, there exists no $R$
which could generate those ensembles
via (\ref{1}) and (\ref{15}).
Nevertheless, both of them still
amount to perfectly well-defined 
ensembles of normalized random 
vectors.

In other words, the first main advantage 
of the ensemble $E$ is that it can be 
easily implemented numerically.
A second advantage is that it is also 
relatively convenient (compared with the 
other two ensembles) for analytical
investigations, as demonstrated with our 
present paper.

The main virtue of the $S$ ensemble
is the conceptual appeal of its
``clean'' and ``natural'' definition
(see below (\ref{m1})).

The main advantage of the GAP ensemble is 
that it often admits exact analytical
statements (in contrast to bounds or
large $N$ asymptotics).

For the rest, all three ensemble result in
practically indistinguishable expectation 
value statistics (dynamical typicality 
properties etc.) for sufficiently large 
dimensions $N$.

Overall, the main conclusion of this section
may be summarized as follows:
The conceptually most ``natural'' ensemble $S$, 
realizing with equal probability all pure states 
$|\psi\rangle$ with $K$ (strictly) preset expectation 
values $\langle\psi |A_k| \psi\rangle$,
is hard to access numerically as well
as analytically.
On the other hand, the essential properties
of that ensemble $S$ are practically
perfectly imitated by the much more
convenient ensemble $E$, which is
readily accessible via 
(\ref{1}), (\ref{15}),  (\ref{29}), (\ref{n7}).
Most importantly, demonstrating dynamical 
typicality properties {\em directly} for the $S$ 
ensemble is expected to be much more difficult
than along our present, ``indirect'' approach
via the $E$ ensemble.

\section{Typicality of equilibration}
\label{s9}
In this section, we focus on isolated systems,
described by a time independent Hamiltonian $H$
with eigenvalues $E_n$ and eigenvectors 
$|\chi_n\rangle$
(see also beginning 
of Sec. \ref{s2} and Sec. \ref{s5}).
At time $t=0$, initial states
$|\psi(0)\rangle = |\psi\rangle$ are
randomly sampled according to (\ref{15}) 
or (\ref{n1}), and then evolve in time
as $|\psi(t)\rangle=U_t|\psi\rangle$
with propagator $U_t=\exp\{-iHt/\hbar\}$,
see also above Eq. (\ref{13}).
Finally, we recall that the dynamical
typicality property (\ref{17})
will be fulfilled in very good 
approximation for most initial states
$|\psi\rangle$, provided $\rho$ 
in (\ref{16}) satisfies (\ref{10}),
or equivalently (\ref{21}).

Following Refs. \cite{lin09,equil},
we say that the system exhibits 
equilibration if the expectation value
$\langle \psi(t)|O|\psi(t)\rangle$
remains very close to a constant value 
for the vast majority of all sufficiently 
large times $t$, i.e., 
after initial transients 
(relaxation processes) 
have died out.
(Note that a small fraction of exceptional 
times $t$ is unavoidable due to quantum
revival effects.)
If this constant long-time value is 
furthermore (approximately) equal
to the pertinent microcanonical 
expectation value, as predicted by
textbook statistical mechanics,
then we say that the system 
exhibits thermalization 
\cite{lin09,equil}.
In the following, we restrict ourselves 
to the question of equilibration,
while the issue of thermalization is 
not pursued any further.

According to (\ref{17}), most initial
states $|\psi\rangle$ will exhibit 
equilibration if the mixed state
$\rho(t)$ exhibits equilibration.
The latter has been demonstrated under 
quite weak 
conditions on the spectrum
of $H$ and on the initial state 
$\rho(0)=\rho$ in Refs. 
\cite{lin09,equil}:
Essentially, it is sufficient that
the energy differences $E_m-E_n$
do not coincide for too many
index pairs with $m\not= n$, 
which is the case for any generic 
Hamiltonian $H$ \cite{lin09,equil}, 
and that all energy
levels are weakly populated, i.e.,
\begin{eqnarray}
\max_n\langle \chi_n|\rho|\chi_n\rangle
\ll 1
\ .
\label{x1}
\end{eqnarray}
Observing that the left hand side
of (\ref{x1}) is upper bounded by the
maximal eigenvalue of $\rho$ 
in (\ref{18a}), which in turn is
given by $p_{max}$ in (\ref{20}),
one readily sees that our previously 
established precondition for dynamical
typicality
from (\ref{21}) is at the same time
sufficient to guarantee
equilibration.

In summary, if the system under
consideration is isolated, exhibits 
a generic energy spectrum, and 
exhibits dynamical typicality,
then most initial states must
also exhibit equilibration.

\section{Practical considerations}
\label{s10}
To explicitly determine the density operator
$\rho$ in (\ref{n7}) and the concomitant $R$ 
in (\ref{29}) by solving (\ref{34}) for a given
set of $\alpha_k$'s may be quite cumbersome
in practice (e.g., by numerical means).
In fact, already in order to define the 
pertinent energy shell in Sec. \ref{s71},
one usually cannot avoid to diagonalize 
the Hamiltonian.

A radically different approach is to start out from
some ``reasonable ansatz'' for $R$ in 
accordance with (\ref{29}) and (\ref{n7}).
In principle, one could then determine the
corresponding $\alpha_k$ values in 
(\ref{34}) {\em a posteriori}, but in 
practice some other way of justifying
the proposed ansatz may 
even be more appropriate.
Also the need to work with an energy shell 
may be circumvented due to the following
observation:
The actually relevant relations
-- for instance (\ref{1})-(\ref{23}), (\ref{34}), and 
(\ref{n7}) -- remain well-defined even
when working with the full Hilbert space 
of the considered model (see also Sec. \ref{s4}),
or, formally speaking, with the ``largest possible 
energy shell'' (its dimension may in principle
be infinite, but numerically it will still be finite 
from the outset).
Yet, a relatively ``sharp''  energy of the (typical) 
initial states $|\psi\rangle$ may be achieved, for
instance, by setting $A_1=H$ and $A_2=H^2$
in (\ref{34}) and choosing some suitable
values of $\alpha_1$ and $\alpha_2$.
Within our present approach, this 
amounts to evaluating the quantities
\begin{eqnarray}
\bar H & := & \tr\{\rho H\}
\ ,
\label{p1}
\\
\delta\! \bar H^2 & := & \tr\{\rho (H-\bar H)^2\}
=\tr\{\rho H^2\} - \bar H^2
\label{p2}
\end{eqnarray}
by means of the proposed ansatz for $\rho$ and
then to decide, whether such an energy distribution
(``washed out energy window'')
is still acceptable to describe 
the problem at hand
(note the similarities and differences 
compared to (\ref{4}) and (\ref{5})).

In order to guarantee dynamical typicality properties,
also (\ref{10}) must be verified by the proposed 
ansatz $\rho$.
As a consequence, instead of evaluating
(\ref{34}), (\ref{p1}), and (\ref{p2}) by means
of $\rho$, the corresponding values 
of $\alpha_k$, $\bar H$, and $\delta\! \bar H^2$,
which this ansatz entails, can
also be approximated very well by 
means of a single (supposedly typical) 
pure state $|\psi\rangle$.

A simple example (see also Ref. \cite{ste14})
is to choose $R$ in (\ref{29})
proportional to $\exp\{-(H-U)^2/4\sigma^2\}$
and hence $\rho$ in (\ref{18a}) as
\begin{eqnarray}
\rho 
& = & 
\frac{1}{Y}\exp\left\{-\frac{(H-U)^2}{2\sigma^2}\right\}
=
\sum_{n=1}^N p_n
\, |n\rangle\langle n|
\ ,
\label{p3}
\\
p_n & := & 
\frac{1}{Y}\exp\left\{-\frac{(E_n-U)^2}{2\sigma^2}\right\}
\ ,
\label{p4}
\end{eqnarray}
where $U$ and $\sigma$ are real parameters,
$Y$ is a normalization constant,
$E_n$ and $|n\rangle$ are the eigenvalues 
and eigenvectors of $H$, and (\ref{30}) 
has been exploited in (\ref{p3}).
Assuming that $\sigma$ is much larger than the 
typical distance between neighboring 
energy levels $E_n$, and that the mean level 
spacing can be approximated by a 
constant value $\Delta$
within a neighborhood of $U$ extending over 
several $\sigma$'s,
sums over $n$ can be readily approximated 
by integrals in (\ref{p1})-(\ref{p3}), yielding
\begin{eqnarray}
\bar H & \simeq & U
\ ,
\label{p5}
\\
\delta\!\bar H & \simeq & \sigma
\ ,
\label{p6}
\\
Y & \simeq & \sqrt{2\pi}\, \sigma/\Delta
\ .
\label{p7}
\end{eqnarray}
With (\ref{20}) and (\ref{p4}) we can 
conclude that that $p_{max}\leq 1/Y$.
Since we assumed $\sigma\gg\Delta$
it follows that our ansatz (\ref{p3}) 
satisfies (\ref{21}) and hence (\ref{10}).
Moreover, we see that the location and width of
the ``washed out energy window'' 
in (\ref{p1}), (\ref{p2}) can be controlled
via the parameters $U$ and $\sigma$
in (\ref{p3}).

According to (\ref{n3}), normalized vectors 
$|\psi\rangle$ with equal values of
$\langle \psi|\rho^{-1}|\psi\rangle=\sum_{n=1}^N |\langle\psi|n\rangle|^2/p_n$
are realized with equal probability.
Moreover, by means of (\ref{30}),
our ansatz (\ref{p3}) can be rewritten 
in the form (\ref{n7}) with $K=\infty$ and
\begin{eqnarray}
\lambda_k & := & Y\ \mbox{for} \ k=0,1,2,...
\ ,
\label{p8}
\\
A_k  & := & \frac{1}{(2\sigma^2)^k k!}\, (H-U)^{2k}
\ \mbox{for} \ k=1,2,3,...
\, .
\label{p9}
\end{eqnarray}
Under the same assumption as above (\ref{p5})
it follows with (\ref{34}) that
\begin{eqnarray}
\alpha _k & \simeq &(2k)!/(2^k k!)^2 \simeq 1/\sqrt{\pi k}
\ ,
\label{p10}
\end{eqnarray}
where the last relation is obtained for large $k$
by Stirling's approximation.
In passing we note that $K\ll N$ has been 
assumed in Sec. \ref{s7} for the sole reason 
that otherwise (\ref{34}) may not be solvable
for almost every {\em a priori} fixed set
of $\alpha_k$'s.
In contrast, here $\rho$ is ``given'' and the
corresponding $\alpha_k$'s are determined
via (\ref{34}) {\em a posteriori}, so that
$K\to\infty$ does not give rise 
to any problems.

From the physical viewpoint, (\ref{p3}) 
represents yet another equilibrium 
ensemble of the type discussed
in Sec. \ref{s6}.
It is therefore plausible that the concomitant
expectation values and hence the typicality
properties for this equilibrium ensemble
will be equivalent (practically
equal) to those of, e.g., the microcanonical
ensemble from (\ref{31}) or
the canonical ensemble 
from  (\ref{32}).
However, from the mathematical
viewpoint this equivalence 
of the ensembles is far from
being obvious and in fact is 
strictly speaking even untenable 
for arbitrary observables.
Rather, the equivalence is only
expected to be true for observables
which satisfy the so-called
eigenstate thermalization 
hypothesis (ETH) 
\cite{deu91,sre94,sre96,rig08,ale16,gog16}.

It is straightforward to formally generalized
(\ref{p3}) for non-equilibrium situations,
for instance by choosing \cite{koh15,koh16}
\begin{eqnarray}
\rho & = & Y^{-1}\,\exp\{-S\}
\ ,
\label{p11}
\\
S & := & 
\frac{(H-U)^2}{2\sigma^2}
+
\frac{(G-V)^2}{2\tau^2}
\ ,
\label{p12}
\end{eqnarray}
where $U$, $V$, $\sigma$, $\tau\in\RR$ are parameters 
and $G$ is some observable, by which the initial
disequilibrium can be characterized, and
which in general does not commute with $H$.
As before, such an ansatz is again of the
general form (\ref{n7}) with (\ref{p8}) and
$A_k:=S^k/k!$ for $k=1,2,3,...$,
while the corresponding $\alpha_k$ 
values can be readily determined
{\em a posteriori} via (\ref{34}).
Likewise, (\ref{10}) will be satisfied
whenever $Y \gg 1$.
It is again reasonable to expect that such an ansatz
may effectively generate some ``washed out 
energy window'' and simultaneously some
reasonably sharply defined
non-equilibrium initial conditions 
for the observable $G$.
The actual energy window 
follows again by evaluating (\ref{p1}) and 
(\ref{p2}), and similarly for the initial
distribution of measurement 
outcomes for $G$.
But unlike in the equilibrium case,
such simple relations between the 
pertinent eigenvalues and eigenvectors
as in (\ref{p3}), (\ref{p4}) are no longer
available.
As a consequence, also
(\ref{p5}) and (\ref{p6}) 
will in general 
no longer be fulfilled. However, for a reasonable 
choice of parameters ($\tau$ not too small, $V$ not too far away from the 
equilibrium expectation value of $G$) it appears plausible 
that the implication of  (\ref{p7}), namely that $\rho$ 
complies with  (\ref{10}), remains valid, c.f. below.
In this case typicality is expected, i.e., most pure states 
of the ensembles as given in (\ref{1}) or (\ref{15}) will reliably  
mimic the behavior of $\rho$.
Other than that, and most importantly,
it is now much more difficult to say
which actual (e.g. experimental or 
numerical) situation and which 
observables 
can be faithfully 
approximated in a quantitative way
within our present typicality approach
by means of such an ansatz for
$\rho$.

On the other hand, the numerical advantage
of such an approach is quite remarkable:
Similarly as described at the end of Sec. \ref{s6},
the operator $R$ in (\ref{1}) can be
rewritten due to (\ref{29}) and (\ref{p11})
as $\gamma\exp\{-S/2\}$ with 
$\gamma=Y^{-1/2}$.
Setting temporarily $\gamma=1$, the 
evaluation of $|\phi\rangle$ in (\ref{1})
is possible by means of numerically 
very efficient imaginary time propagation 
techniques
\cite{koh15,koh16,sch18,jin16}, 
yielding the same normalized 
random vector $|\psi\rangle$ 
in (\ref{15}) as one would have 
obtained for $\gamma=Y^{-1/2}$.
But by working with 
$\gamma=1$, the resulting
random number $\langle\phi | \phi\rangle$
will be typically very close to $Y$.
In other words, the same numerically
obtained random vector $|\phi\rangle$
can be used not only to deduce the
usual dynamical typicality properties
but also to check whether the above derived
sufficient condition $Y\gg 1$ for (\ref{10}) is 
fulfilled.

%
\section{summary and conclusion}

We examined ensembles of pure states which are 
distributed in a high-dimensional
Hilbert space such that states 
featuring  equal expectation values for a set of $K$ 
observables $A_k$ occur with equal probability. 
These ensembles are usually not Haar-measure invariant. 
Rather they may be considered as unbiased
ensembles, given an  {\it a priori} knowledge about the 
expectation values of the above observables $A_k$.
Such an  {\it a priori} knowledge is characteristic 
of the initial situation of non-equilibrium experiments 
in which the relaxation of some observables is monitored.
Under rather mild conditions, these ensembles are found 
to exhibit strong typicality properties, i.e., the overwhelming 
majority of individual states from these ensembles feature very 
similar expectation values 
for a largely arbitrary observable
at the initial as well as at any later instance of time.
In particular, this observable may be different from 
the above $A_k$'s or linear combinations thereof.

In other words, we demonstrated that knowing
the initial expectation values of a few observables is sufficient
to produce a very reliable guess about the 
time-dependent expectation values of another, 
possibly very different observable. 
In particular this gives rise to so-called ``dynamical typicality'': 
Knowledge of some initial expectation value  
$\langle\psi|O(0) | \psi\rangle$ suffices to produce a 
very reliable guess on the expectation value of the 
same observable at any later time $\langle\psi|O(t) | \psi\rangle$.

We discussed in detail three ensembles which exhibit
the above ``unbiasedness'' but differ in the actual probability 
with which pure states featuring some common 
expectation values actually occur in the ensemble. 
This leads to mathematically well-defined qualitative 
differences between these ensembles. It is demonstrated, 
however, that the above guess on the expectation 
value of some further observable only varies very 
slightly with the choice of a specific ensemble. 
One of the discussed ensembles may be efficiently 
numerically generated on a computer. 
This allows for the construction of a numerical 
tool with which the behavior of ensembles,
mixed states, etc. may be mimicked by a single, 
properly chosen pure state.

\begin{acknowledgments}
Stimulating discussions with  
Ben N. Balz and Lennart Dabelow
are greatefully acknowledged.
We also thank Shelly Goldstein 
for explaining the derivation 
of Eq. (18a) in \cite{gol06a}.
This work was supported by the 
Deutsche Forschungsgemeinschaft (DFG)
within the Research Unit FOR 2692
under Grants No. 397303734
and 397107022.
\end{acknowledgments}

\appendix
\section{Time-dependent Hamiltonians}
\label{app1}
As mentioned above (\ref{13}), the system
dynamics may be generated in full generality 
by an explicitly time-dependent Hamiltonian.
In case that the two Hilbert spaces
$\hr$ and $\hr\oo$ disagree 
(see Sec. \ref{s4}), certain details
of the so far formalism require
some refinements:

In the case of an explicitly time 
dependent Hamiltonian $H\oo(t)$, 
we still can conclude from Sec. \ref{s4}
that $H(0):=P\oo H\oo(0)P\oo=H\oo(0)$.
Hence, the discussion in Sec. \ref{s2} 
still remains valid as long as only 
the time $t=0$ matters.
The first relevant modification thus
concerns the propagator $U_t\oo$ 
(see above (\ref{13})), which is
{\em a priori} an operator on 
$\hr\oo$.
For time independent Hamiltonians,
one sees that
the relation $U_t:=P\oo U_t\oo P\oo=U_t\oo$ 
holds for all $t$, 
hence the 
transition from $\hr\oo$ to $\hr$
is trivial.
The same no longer applies 
for time-dependent Hamiltonians.
Yet, (\ref{13}) remains valid
if we define $O_t:=P\oo O_t\oo P\oo$,
and replace (\ref{14}) by
\begin{eqnarray}
O_t\oo:=(U_t\oo)^\dagger O\oo U_t\oo \ .
\label{a1}
\end{eqnarray}

Analogously, 
$\rho\oo(t):=(U_t\oo)^\dagger \rho\oo(0)\oo U_t\oo$
is now an operator on $\hr\oo$ and
(\ref{17}) has to be replaced by
\begin{eqnarray}
\langle \psi (t)|O| \psi (t)\rangle
=\tr\oo\{\rho\oo(t)\, O\oo\} \ ,
\label{a2}
\end{eqnarray}
where $\tr\oo$ indicates the
trace in $\hr\oo$.
Note that (\ref{a2}) and (\ref{17}) 
are still equivalent for $t=0$.

In spite of these modifications,
all the dynamical typicality properties
discussed below (\ref{17}) and 
in Sec. \ref{s4}
still  remain valid even for explicitly 
time-dependent Hamiltonians.

\section{Derivation of Equation (\ref{n3})}
\label{app2}
In this Appendix we show that the probability 
distribution of the random vectors in (\ref{n1}) 
is given by (\ref{n3}).

As said below (\ref{1}), the real and imaginary parts 
of the complex numbers $c_n$ are given by independent, 
Gaussian distributed random variables of mean 
zero and variance $1/2$.
Accordingly, the $N$-dimensional complex vectors 
\begin{eqnarray}
{\bf c} & := & (c_1,...,c_N)
\label{b1}
\end{eqnarray}
can and will be interpreted from now on as 
$2N$-dimensional real vectors, 
whose probability distribution is given by
\begin{eqnarray}
w_1({\bf c}) &: = &
\pi^{-N}\,
\exp\left\{-\norm{{\bf c}}^2
\right\}
\ ,
\label{b2}
\\
\norm{\bf c} & := & 
\sqrt{
\mbox{$\sum_{n=1}^N|c_n|^2$}
}
\ .
\label{b3}
\end{eqnarray}

Viewing (\ref{n2}) as a transformation
of random variables,
the connection between the corresponding
probability distributions $w_1({\bf c})$ and
$w({\bf y})$ is given according
to textbook probability theory by
\begin{eqnarray}
w({\bf y}) & := & \int d{\bf c}\, w_1({\bf c})\
\delta({\bf y}-\tilde{\bf y}({\bf c}))
\ ,
\label{b4}
\\
\tilde y_n({\bf c}) & := & \frac{\sqrt{p_n}\,c_n }{\sqrt{\sum_{n=1}^N p_n |c_n|^2}}
\ ,
\label{b5}
\end{eqnarray}
with the following definitions for any ${\bf z}\in\CC^N$:
\begin{eqnarray}
d{\bf z} & := & 
\prod_{n=1}^N d(\mbox{Re} z_n) \  d(\mbox{Im} z_n)
\ ,
\label{b6}
\\
\delta({\bf z}) & := &
\prod_{n=1}^N \delta(\mbox{Re} z_n) \ \delta(\mbox{Im} z_n) 
\ .
\label{b7}
\end{eqnarray}
Going over from the integration variables $c_n$ to $z_n:=\sqrt{p_n}c_n$
yields
\begin{eqnarray}
w({\bf y}) & = & \int d{\bf z}\, w_2({\bf z})\
\delta({\bf y}-{\bf z}/\norm{\bf z})
\ ,
\label{b8}
\\
w_2({\bf z}) & := &
\frac{1}{\pi^N\det(\rho)}\,
\exp\left\{-\sum_{n=1}^N \frac{|z_n|^2}{p_n}\right\}
\ ,
\label{b9}
\end{eqnarray}
where $\det(\rho):=\prod_{n=1}^Np_n$ is the determinant
of $\rho$ from (\ref{18a}).
Multiplying the right hand side of (\ref{b8}) by
$1=\int_0^\infty dr\ \delta({\norm{\bf z}}-r)$ result in
\begin{eqnarray}
w({\bf y}) = \int_0^\infty\!\! 
dr \int d{\bf z}\, w_2({\bf z})\
\delta({\bf y}-{\bf z}/r)\ \delta({\norm{\bf z}}-r)
\ .\ \ \ \ 
\label{b10}
\end{eqnarray}
Going over from the integration variables $z_n$
to $v_n:=z_n/r$ implies
\begin{eqnarray}
w({\bf y}) & = & \int_0^\infty\!\! dr \int d{\bf v}\ r^{2N}\,
w_2(r{\bf v})\ \delta({\bf y}-{\bf v})\ \delta(r{\norm{\bf v}}-r)
\nonumber
\\
& = & 
\int_0^\infty dr \, r^{2N}\,
w_2(r{\bf y})\ \delta(r({\norm{\bf y}}-1))
\ .
\label{b11}
\end{eqnarray}
Defining the Heaviside step function for any $x\in\RR$
as $\Theta(x):=\int_{-\infty}^x dy \,\delta (y)$,
it follows that $\Theta'(x)=\delta(x)$ and
$\Theta(rx)=\Theta(x)$ for any $r>0$.
One thus can conclude that
$\delta(x)=d \Theta(x)/dx=d \Theta(rx)/dx=
r\Theta'(rx)=r\delta(rx)$
and therefore 
$\delta(rx)=\delta(x)/r$.
Exploiting the latter identity and (\ref{b9}),
we can rewrite (\ref{b11}) as
\begin{eqnarray}
w({\bf y}) & = & 
\frac{\delta({\norm{\bf y}}-1)}{\pi^N\det(\rho)}
\int_0^\infty dr \, r^{2N-1}\,
e^{-r^2 g({\bf y})}
\ ,
\label{b12}
\\
g({\bf y}) & := & \sum_{n=1}^N \frac{|y_n|^2}{p_n}
\ .
\label{b13}
\end{eqnarray}
Going over from the integration
variable $r$ to $t:=g({\bf y})r^2$ and exploiting 
the common properties of the gamma function,
one finds that
\begin{eqnarray}
w({\bf y}) & = & 
\frac{\delta({\norm{\bf y}}-1)}{\pi^N\det(\rho)}
\frac{(N-1)!}{2\,g({\bf y})^N}
\ .
\label{b14}
\end{eqnarray}

Next we observe that (\ref{n1}) represents a 
one-to-one correspondence between any 
possible vector $|\psi\rangle\in\hr$ and 
any possible vector ${\bf y}$ in $\CC^N$,
or equivalently in $\RR^{2N}$.
Accordingly, and as said already below (\ref{n1}), 
the probability distribution
$w(|\psi\rangle)$ of the vectors 
$|\psi\rangle\in\hr$ is by definition
given by the probability distribution 
of the corresponding 
random numbers $y_n$  in (\ref{n1}),
i.e., by $w({\bf y})$ from (\ref{b14}).
Moreover, $\norm{\bf y}$ can be readily 
identified with $\norm{\psi}$.
Likewise, one can rewrite (\ref{b13}) by means of
(\ref{30a}) and (\ref{n1}) as
\begin{eqnarray}
g({\bf y}) =\langle\psi|\rho^{-1}|\psi\rangle
\ .
\label{b15}
\end{eqnarray}
Taking all this onto account in (\ref{b14}),
one finally recovers (\ref{n3}).

\section{Derivation of Equation (\ref{n7})}
\label{app3}
In this Appendix we show that if there exists 
as function $h:\RR^K\to\RR$ which fulfills
(\ref{n5}) for all normalized 
vectors $|\psi\rangle\in\hr$, then
$\rho$ must be of the form
Eq. (\ref{n7}) (under the 
tacit extra assumption that $h$ is twice 
continuously differentiable).

Clearly, (\ref{n5}) is satisfied for all normalized 
vectors $|\psi\rangle\in\hr$ if and only if
\begin{eqnarray}
\langle\phi|\rho^{-1}|\phi\rangle 
=
\norm{\phi}^2\,
h(a_\phi^1,...,a_\phi^K)
\label{c1}
\end{eqnarray}
with $a_\phi^k:=\langle\phi |A_k|\phi\rangle/\norm{\phi}^2$
and $\norm{\phi}:=\langle\phi|\phi\rangle^{1/2}$
is satisfied for all vectors 
$|\phi\rangle\in\hr$ of non-zero but 
otherwise arbitrary norm.

Rewriting  (\ref{30a}) by means of the 
abbreviation $q_n:=1/p_n$ as
\begin{eqnarray}
\rho^{-1}=\sum_{n=1}^N q_n\, |n\rangle\langle n|
\ ,
\label{c2}
\end{eqnarray}
we now focus in (\ref{c1}) on the special choice
\begin{eqnarray}
|\phi\rangle = |m\rangle+\sqrt{\epsilon}\, z\, |n\rangle
\label{c3}
\end{eqnarray}
for arbitrary $m\not=n$, $z\in\CC$, and $\epsilon\in\RR_0^+$
(later we will set $\epsilon=1$ and restrict ourselves to small $|z|$).
Introducing (\ref{c2}) and (\ref{c3}) into (\ref{c1}) 
yields
\begin{eqnarray}
& & 
q_m+\epsilon z^2 q_n
= (1+\epsilon z^2)\, h(x_1,...,x_K)
\ ,
\label{c4}
\\
& & 
x_k 
:=
\frac{
A^k_{mm}+\sqrt{\epsilon}Z^k_{mn}+\epsilon z^2A^k_{nn}
}{1+\epsilon z^2}
\ ,
\label{c5}
\\
& &
A^k_{mn}:=\langle m|A_k|n\rangle
\ ,
\label{c5a}
\\
& & 
Z_{mn}^k:=z A^k_{mn}+(z A^k_{mn})^\ast
=
2\, \mbox{Re}(zA^k_{mn})
\ .
\label{c6}
\end{eqnarray}
Differentiating (\ref{c4}) with respect to $\epsilon$
and then setting $\epsilon=1$ yields
\begin{eqnarray}
\!\!\!\!\!\!\!\!\!\!
& & z^2 q_n-z^2 h(x_1,...,x_K)
=
\sum_{k=1}^K y_k\, \frac{\partial h(x_1,...,x_K)}{\partial x_k}
\ , \ \ \ \ 
\label{c7}
\\
\!\!\!\!\!\!\!\!\!\!
& & 
y_k 
:=
\frac{Z^k_{mn}}{2}
+z^2
\left(A_{nn}-\frac{
A^k_{mm}+Z^k_{mn}+z^2A^k_{nn}
}{1+z^2}\right)
\ . \ \ \ \ \ 
\label{c8}
\end{eqnarray}
Upon eliminating $h(x_1,...,x_K)$ in (\ref{c7})
by means of 
(\ref{c4}) (with $\epsilon=1$) one obtains
\begin{eqnarray}
\frac{z^2}{1+z^2}\, (q_n-q_m)
=
\sum_{k=1}^K y_k\, \frac{\partial h(x_1,...,x_K)}{\partial x_k}
\ . \ \ \ 
\label{c9}
\end{eqnarray}

Focusing on small $|z|$ and observing that $Z_{mn}^k$
in (\ref{c6}) is of first order in $z$, it follows from
(\ref{c5}) and (\ref{c8}) that
\begin{eqnarray}
& & 
y_k\, \frac{\partial h(x_1,...,x_K)}{\partial x_k}
=
\left(
\frac{Z^k_{mn}}{2}+z^2(A^k_{nn}-A^k_{mm})
\right)
h_k^m
\nonumber
\\
& & 
\qquad\qquad\qquad\ \ 
+\frac{1}{2}\sum_{j=1}^K
Z^j_{mn} Z^k_{mn} h_{jk}^m+\ord(z^3)
\ ,
\label{c10}
\\
& & 
h_{k}^m := 
\frac{\partial h(A^1_{mm},...,A^K_{mm})}{\partial x_k}
\ ,
\label{c11}
\\
& & 
h_{jk}^m := 
\frac{\partial^2 h(A^1_{mm},...,A^K_{mm})}{\partial x_j\partial x_k}
\ .
\label{c12}
\end{eqnarray}
Since $h$ is a real valued function, also its derivatives in
(\ref{c11}) and (\ref{c12}) are real numbers.

By collecting terms of first order in $z$ one can infer
from (\ref{c9}) and (\ref{c10}) that
\begin{eqnarray}
& & 
0= \sum_{k=1}^K Z^k_{mn} h_k^m = z G_{mn}^m+(z G_{mn}^m)^\ast
\ , 
\label{c13}
\\
& & G^l_{mn} := \sum_{k=1}^K A_{mn}^k h_k^l 
\ ,
\label{c14}
\end{eqnarray}
where we exploited (\ref{c6}) on the right hand side
of (\ref{c13}).
Since $z\in\CC$ in (\ref{c13}) is  
still arbitrary, it follows that
\begin{eqnarray}
G_{mn}^m=0
\ ,
\label{c15}
\end{eqnarray}
where, as said below (\ref{c3}), the indices $m$ and $n$
must be different but otherwise may be arbitrary.

Analogously, by collecting terms of second order 
in $z$ one can infer from (\ref{c9}) 
and (\ref{c10}) that
\begin{eqnarray}
\!\!
z^2 (q_n-q_m)
=
z^2(G^m_{nn}-G^m_{mm}) 
+
\!\!\sum_{j,k=1}^K\! \frac{Z^j_{mn}Z^k_{mn}\, h_{jk}^m}{2}
\, . \ \ \ \ 
\label{c16}
\end{eqnarray}
One readily confirms that apart from $z$, all other 
terms in (\ref{c16}) are real numbers.
Since $z\in\CC$ in (\ref{c16}) is \
still arbitrary, one can conclude that
\begin{eqnarray}
& & 
q_n-q_m = G^m_{nn}-G^m_{mm} 
\label{c17}
\end{eqnarray}
for arbitrary $m\not=n$.
Obviously, this result also remains valid for $m=n$.

Considering $m$ as arbitrary but fixed and
$n$ as variable, we can rewrite
(\ref{c17}) with (\ref{c14}) as
\begin{eqnarray}
q_n & = & \lambda_0^m+\sum_{k=1}^K\lambda_k^m A^{k}_{nn}
\ ,
\label{c18}
\\
\lambda_0^m & := & q_m-G_{mm}^m
\ ,
\label{c19}
\\
\lambda_k^m & := & h_k^m
\ .
\label{c20}
\end{eqnarray}
Note that all quantities in (\ref{c19})
are real numbers.
Considering $m$ as arbitrary but fixed,
(\ref{c18}) amounts to a linear relation
between the $q_n$ and the $A_{nn}^k$ for all
$n=1,...,N$.
If we choose some different value $m$
as our fixed index, the same relation must
again be true for all $n=1,...,N$.
Since the left hand side of (\ref{c18}) is
independent of $m$ and since $K\leq N$
(see main text), this is only possible
if the coefficients $\lambda_0^m,...,\lambda_K^m$
are actually the same for any choice of $m$.
In other words, the coefficients are independent of $m$
and hence their upper index $m$ can be omitted.
Accordingly, (\ref{c18}) takes the form
\begin{eqnarray}
q_n & = & \lambda_0+\sum_{k=1}^K\lambda_k A^{k}_{nn}
\label{c21}
\end{eqnarray}
for all $n=1,...,N$.
Likewise, (\ref{c14}) can be rewritten with
(\ref{c20}) as
\begin{eqnarray}
G^l_{mn} = \sum_{k=1}^K \lambda_k A_{mn}^k 
\ ,
\label{c22}
\end{eqnarray}
independent of $l$.
Finally, (\ref{c15}) thus amounts to
\begin{eqnarray}
\sum_{k=1}^K \lambda_k A_{mn}^k = 0
\label{c23}
\end{eqnarray}
for all $m\not=n$.

Taking into account  (\ref{c2}), (\ref{c5a}), (\ref{c21}), and
(\ref{c23}), one readily confirms that
\begin{eqnarray}
\langle m| \rho^{-1}|n\rangle 
= \lambda_0\, \delta_{mn} + \sum_{k=1}^K \lambda_k \langle m| A_k|n\rangle
\ \ 
\label{c24}
\end{eqnarray}
for all $m,n$.
Consequently,
\begin{eqnarray}
\rho^{-1}:= \lambda_0 \id + \sum_{k=1}^K \lambda_k A_k
\ ,
\label{c25}
\end{eqnarray}
and thus Eq. (\ref{n7}) from the 
main text follows.

Introducing (\ref{n7}) into (\ref{n5}) implies that
\begin{eqnarray}
h(v_1,...,v_K) = 
\lambda_0 + \sum_{k=1}^K \lambda_k v_k
\ .
\label{c26}
\end{eqnarray}
In turn, it is straightforward to verify that 
this function with arbitrary real coefficients
$\lambda_0,...,\lambda_K$
fulfills all the equations and properties
which we encountered in the course of the
above calculations.
Our considerations also show that 
only functions $h$ of the form (\ref{c26})
satisfy all those equations and properties.

\section{Reducing the number of equations}
\label{app4}
In view of (\ref{n7}) it follows
that
\begin{eqnarray}
\tr\left\{\rho \,
\left[\lambda_0\, \id + \sum_{k=1}^K\lambda_k\, A_k\right]
\right\} =N
\label{103}
\end{eqnarray}
and hence
\begin{eqnarray}
\lambda_0
\tr\{\rho\} +\sum_{k=1}^K\lambda_k\, \tr\{\rho A_k\} = N
\ .
\label{104}
\end{eqnarray}
Together with conditions (\ref{3}) and (\ref{34})
we can conclude that
\begin{eqnarray}
\lambda_0=N-\sum_{k=1}^K\lambda_k\, 
\ak_k
\ .
\label{105}
\end{eqnarray}
In other words, among the $K+1$ conditions
(\ref{3}) and (\ref{34}) we can choose 
an arbitrary one and replace it by (\ref{105}).

Defining $y_k:=-\lambda_k/N$ and
introducing (\ref{105}) into (\ref{n7})
yields
\begin{eqnarray}
\rho & = & \frac{1}{N}\,
\left[\id - 
\sum_{k=1}^K y_k\, (A_k-\ak_k)\right]^{-1}
\ ,
\label{106}
\end{eqnarray}
where the $K$ parameters $y_k$ are fixed
by the $K+1$ conditions (\ref{3}) and 
(\ref{34}), among which one will be 
automatically fulfilled if all the $K$ 
others are fulfilled.
Moreover, the $y_k$'s must be so
that (\ref{106}) is a strictly positive
definite operator (cf. (\ref{91})).

A more detailed exploration of the
special case $K=1$ has been
carried out in \cite{rei18}.
In particular, such a closer inspection shows 
that the minus sign in the above definition 
of the $y_k$ is convenient and natural.



\begin{thebibliography}{99}

\bibitem{eng01}
A. Engel and C. Van den Broeck,
{\em Statistical Mechanics of Learning}
(Cambridge University Press, 2001)

\bibitem{gol06}
S. Goldstein, J. L. Lebowitz, R. Tumulka, and N. Zanghi, 
Phys. Rev. Lett. {\bf 96}, 050403 (2006)

\bibitem{pop06}
S. Popescu, A. J. Short, and A. Winter, 
Nature Phys. {\bf 2}, 754 (2006)

\bibitem{llo88}
S. Lloyd, Ph.D. Thesis, 
The Rockefeller University (1988),
Chapter 3, arXiv:1307.0378

\bibitem{bar09}
C. Bartsch and J. Gemmer, 
Phys. Rev. Lett. {\bf 102}, 110403 (2009)

\bibitem{sug07}
A. Sugita, 
Nonlinear Phenom. Complex Syst. {\bf 10}, 192-195 (2007)

\bibitem{rei07}
P. Reimann,  Phys. Rev. Lett. {\bf 99}, 160404 (2007)

\bibitem{gem09}
J. Gemmer, M. Michel, and G. Mahler, 
{\em Quantum Thermodynamics} 
(2nd edition, Springer, Berlin, Heidelberg, 2009)

\bibitem{sug12}
S. Sugiura and A. Shimizu, Phys. Rev. Lett. {\bf 108}, 
240401 (2012)

\bibitem{tas16}
H. Tasaki,
J. Stat. Phys. {\bf 163}, 937 (2016)

\bibitem{bal18}
B. N. Balz, J. Richter,
J. Gemmer, R. Steinigeweg,
and P. Reimann,
Dynamical typicality for initial states with 
a preset measurement statistics of
several commuting observables, in
{\em Thermodynamics in the Quantum Regime;
Fundamental Aspects and New Directions},
edited by F. Binder, L. A. Correa, 
C. Gogolin, J. Anders, and G. Adesso,
Springer, Berlin, Heidelberg, 2018.

\bibitem{rei18}
P. Reimann, 
Phys. Rev. E {\bf 97}, 062129 (2018)

\bibitem{lin09}
N. Linden, S. Popescu, A. J. Short, and A. Winter, 
Phys. Rev.  E {\bf 79}, 061103 (2009)

\bibitem{mal14}
A. S. L. Malabarba, L. P. Garcia-Pintos,
N. Linden, T. C. Farrelly, and A. J. Short, 
Phys. Rev. E 90, 012121 (2014)

\bibitem{rei15}
P. Reimann, Phys. Rev. Lett. {\bf 115}, 010403 (2015)

\bibitem{bar17}
C. Bartsch and J. Gemmer,
Europhys. Lett. {\bf 118}, 10006  (2017)


\bibitem{sug13}
S. Sugiura and A. Shimizu,
Phys. Rev. Lett. {\bf 111}, 010401 (2013)

\bibitem{ham00}
A. Hams and H. De Raedt,
Phys. Rev. E {\bf 62}, 4365 (2000)

\bibitem{r2}
R. Alben, M. Blume, H. Krakauer, and L. Schwartz,
Phys. Rev B {\bf 12}, 4090 (1975);
P. de Vries and H. De Raedt, 
Phys. Rev. B {\bf 47}, 7929 (1993);
T. Iitaka, S. Nomura, H. Hirayama, X. Zhao, Y. Aoyagi, 
and T. Sugano, Phys. Rev. E {\bf 56}, 1222 (1997)

\bibitem{hyu14}
M. Hyuga, S. Sugiura, K. Sakai, and A. Shimizu,
Phys. Rev. B {\bf 90}, 121110(R) (2014)

\bibitem{rei19}
P. Reimann and J. Gemmer,
Phys. Rev. E {\bf 99}, 012125 (2019)

\bibitem{fine09}
B. V. Fine,
Phys. Rev. E {\bf 80}, 051130 (2009)

\bibitem{mueller11}
M. P. M\"uller, D. Gross, and J. Eisert,
Commun. Math. Phys. {\bf 303}, 785 (2011)

\bibitem{gol06a} 
S. Goldstein,  J. L. Lebowitz, R. Tumulka, and N. Zanghi, 
J. Stat. Phys. {\bf  125}, 1197 (2006)

\bibitem{rei08}
P. Reimann,
J. Stat. Phys.  {\bf 132}, 921 (2008)

\bibitem{equil}
P. Reimann, 
Phys. Rev. Lett. {\bf 101}, 190403 (2008);
A. J. Short, 
New J. Phys. {\bf 13}, 053009  (2011);
P. Reimann and M. Kastner, 
New J. Phys. {\bf 14}, 043020 (2012);
A. J. Short and T. C. Farrelly, 
New J. Phys. {\bf 14}, 013063 (2012);
B. N. Balz and P. Reimann,
Phys Rev. E {\bf 93}, 062107 (2016)

\bibitem{ste14}
R. Steinigeweg, A. Khodja, H. Niemeyer, 
C. Gogolin, and J. Gemmer,
Phys. Rev. Lett. {\bf 112}, 130403 (2014)

\bibitem{deu91}
J. M. Deutsch,
Phys. Rev. A {\bf 43}, 2046 (1991)

\bibitem{sre94}
M. Srednicki, 
Phys. Rev. E {\bf 50}, 888 (1994)

\bibitem{sre96}
M. Srednicki, 
J. Phys. A {\bf 29}, L75 (1996)

\bibitem{rig08}
M. Rigol, V. Dunjko, and M. Olshanii, 
Nature (London) {\bf 452}, 854 (2008)

\bibitem{ale16}
L. D'Alessio, Y. Kafri, A. Polkovnikov, and M. Rigol,
Adv. Phys. {\bf 65}, 239 (2016)

\bibitem{gog16}
C. Gogolin and J. Eisert,
Rep. Prog. Phys. {\bf 79}, 056001 (2016)

\bibitem{koh15}
A. Khodja, R. Steinigeweg, and J. Gemmer,
Phys. Rev. E {\bf 91}, 012120 (2015)

\bibitem{koh16}
A. Khodja, D. Schmidtke, and J. Gemmer,
Phys. Rev. E {\bf 93}, 042101 (2016)

\bibitem{sch18}
D. Schmidtke, L. Knipschild, M. Campisi, R. Steinigeweg, and J. Gemmer,
Phys. Rev. E  {\bf 98}, 012123  (2018)

\bibitem{jin16}
F. Jin, R. Steinigeweg, H. De Raedt, K. Michielsen, M. Campisi, and J. Gemmer,
Phys. Rev. E, {\bf 94}, 012125, (2016)

\end{thebibliography}
\end{document}